\documentclass{llncs}
\usepackage[german,english]{babel}
\usepackage{times,theorem}
\usepackage{latexsym,color,graphics}
\usepackage{url}
\usepackage{epsfig}
\usepackage{amssymb}
\usepackage{amsmath}
\usepackage{amsfonts}
\usepackage{diagrams11}
\diagramstyle[tight,centredisplay%
,dpi=600%              for more modern laser-printers
,UglyObsolete%         formerly called ``noPostScript''
,heads=LaTeX%
]
\begingroup
\catcode`\~=11
\gdef\urltilde{\lower 0.6ex\hbox{~}}
\endgroup

\newcommand{\A}{\mathcal{A}} 
 \newcommand{\D}{\mathcal{D}}
\newcommand{\E}{\mathcal{E}} 
 
\newcommand{\I}{\mathcal{I}} 
 \renewcommand{\L}{\mathcal{L}}
\newcommand{\M}{\mathcal{M}} \newcommand{\N}{\mathcal{N}}
 \renewcommand{\P}{\mathcal{P}}
 
\renewcommand{\S}{\mathcal{S}} \newcommand{\T}{\mathcal{T}}
 \newcommand{\V}{\mathcal{V}}
\newcommand{\W}{\mathcal{W}}

\title{Intensional RDB Manifesto: a Unifying NewSQL Model for Flexible Big Data}
\author{Zoran Majki\'c}
\authorrunning{Zoran Majki\'c}
\institute{ISRST, Tallahassee, FL, USA\\
\email{majk.1234@yahoo.com}\\ http://zoranmajkic.webs.com/}
\authorrunning{Zoran Majki\'c}
\newtheorem{propo}{Proposition}
\newcount\pdfoutput
\begin{document}

% \firstpage{1}

\maketitle

\begin{abstract}
In this paper we present a new family of Intensional RDBs (IRDBs)
which extends the traditional RDBs with the Big Data and flexible
and 'Open schema' features, able to preserve the user-defined
relational database schemas and all preexisting user's applications
containing the SQL statements for a deployment of such a relational
data. The standard RDB data is parsed into an internal vector
key/value relation, so that we obtain a column representation of
data used in Big Data applications, covering the key/value and
column-based Big Data applications as well, into a unifying RDB
framework. We define a query rewriting algorithm, based on the GAV
Data Integration methods,  so that each user-defined SQL query is
rewritten into a SQL query over this vector relation, and hence the
user-defined standard RDB schema is maintained as an empty global
schema  for the RDB schema modeling of data and as the SQL interface
to stored vector relation. Such an IRDB architecture is adequate for
the massive migrations from the existing slow RDBMSs into this new
family of fast IRDBMSs  by offering a Big Data and new flexible
schema features as well.
\end{abstract}

%\begin{keywords}
%Keywords: Inovative Web-based Applications, Semantic Web,Web Query
%Language,Web Information retrieval.
%\end{keywords}

%\newpage
%----------------------------------------------------------------------
% SECTION I: Introduction
%----------------------------------------------------------------------
\section{Introduction}
The term NoSQL was picked out in 2009 and used for conferences of
advocates of non-relational databases. In an article of the
Computerworld magazine \cite{Com09a}, June 2009, dedicated to the
NoSQL meet-up in San Francisco is reported the following: "NoSQLers
came to share how they had overthrown the tyranny of slow, expensive
relational databases in favor of more efficient and cheaper ways of
managing data". In this article, the Computerworld summarizes the
following reasons:
\begin{itemize}
  \item \emph{High Throughput}. The NoSQL databases provide a significantly
  higher data throuhput than traditional RDBMSs.
  \item \emph{Horizontal Scalability}. In contrast to RDBMSs most NoSQL
  databases are designed to scale well in the horizontal direction
  and not rely on highly available hardware.
  \item \emph{Cost Setting and Complexity of  Database Clusters}. NoSQL does not need
  the complexity and cost of sharding which involves cutting up
  databases into multiple tables to run on large clusters or grids.
  \item \emph{"One size fits all" \cite{StUg05} Database Thinking Is Wrong}.
  The thinks that the realization and the search for alternatives
  towards traditional RDBMs can be explained by the following two
  major trends: The continuous growth of data volumes and the
  growing need to process larger amounts of data in shorter time.
  \item  \emph{Requirement of Cloud Computing}. Are mentioned two
  major requirements: High until almost ultimate scalability
  (especially in the horizontal direction) and low administration
  overhead. Developed cloud Amazon's SimpleDB can store large
  collections of items which themselves are hashtables containing
  attributes that consist of key-value pairs.
  \item \emph{Avoidance of Unneeded Complexity}.  The reach feature
  set and the ACID properties implemented by RDBMSs might be more
  than necessary for particular applications. There are different
  scenarios where applications would be willing to compromise
  reliability for better performances.
  \end{itemize}
  Moreover, the NoSQL movements advocate that relational fit well
  for data that is \emph{rigidly structured} with relations and are
  designated for central deployments with single, large high-end machines, and not for distribution. Often
  they emphasize that SQL queries are expressed in a sophisticated
  language. But usually they do not tell that also the NoSQL
  databases often need the sophisticated languages (as
  object-oriented databases, or Graph-based databases) as well.
  Moreover, they do not say that SQL and RDB are based on sound
  logical framework (a subset of the First-Order Logic (FOL)
  language) and hence it is not a procedural language, but a
  higher level declarative language able to specify "what" we need
  instead of "how" to obtain what we need. Thus, from the point of view of the development
  of computer science, instead to go in the direction of the
  logically higher levels of knowledge representation and query
  languages, more appropriated to the human understanding, they propose the old
  technics as key-value representations or the simpler forms of
  object-oriented representations and  their relative procedural query languages.\\
  They jumped into the past instead to jump in the future of the social and
  scientific human development.  It happened because the current RDBMSs
  were  obsolete and not ready to accept the new social-network Web applications in the
  last 10 years, so that the isolated groups of developers of these ad-hoc systems (e.g., Google, Amazon, LinkedIn, Facebook, etc..)
  could use only the ready old-known technics and development instruments in order
  to satisfy the highly urgent business market requirements.
  From the academic research side, instead, most of the work has been
  done in Sematic Web "industrial-funded" programs (e.g., the European IST projects) by considering the new knowledge and
  reasoning logic systems, whose impact to the existing RDB
  applications framework would be very hard, with difficult migration and
  implementation in these new semantics systems (it would need one or more
  decade of time). Instead, we needed a more fundamental theoretical
  research  for the significative technological advances and evolutions of the
  existing RDB engine. Thus, the core RDB technology was in some way
  "abandoned" from both major development initiatives in the last 20
  years. Nobody  probably wanted to consider the most natural evolution of the
  RDBMSs and its FOL and SQL query framework, and the database
  industry tried only to cover "with pieces" and "adding" the new emergent
  customer's
  necessities, without a strong investment and the necessary efforts for the
  complete revision of their old System R based RDB engines of the 1970s.
  The world's economical crisis form 2007  did not help for such an effort. \\
  However, from the technical point of view, it is clear that if we would come back to make the application
  programs in Assembler, we probably will obtain better computational
   performances for some algorithms than with more powerful programming languages,
   but it is justifiable when we write the system infrastructures and parsers,
   and not when we have to develop the legacy software for user's requirements.
   Analogously, we may provide the BD infrastructure and physical level in a form of
   simpler structures, adequate to support the distributive and massive BigData query
   computations, by preserving the logically higher level interface to
   customer's applications. That is, it is possible to preserve the RDB interface to data, with
   SQL query languages  for the programmers of the software applications,
   with the "physical" parsing of data in more simple structures, able to deal with
   Big Data scalability in a high distributive computation framework. \\
   The first step to maintain the logical declarative (non-procedural) SQL query language
   level, is done by the group (M.I.T. and Microsoft) and in widely adopted paper "The End of
   an Architectural Era" (cf. \cite{SMAH07} Michael Stonebraker et all.)
    where the authors come to the conclusion "that the current RDBMS code lines, while
   attempting to be a "one size fits all" solution, in fact excel at
   nothing". At first, Stonebraker et all. argue that RDBMSs have
   been architected more than 25 years ago when the hardware
   characteristics, user requirements and database markets where
   very different from those today. The resulting revision of
   traditional RDBMSs is provided by developing H-store (M.I.T., Brown and Yale University), a next generation OLTP systems that operates on
   distributed clusters of shared-nothing machines where the data
   resides entirely in main memory, so that it was shown to
   significantly outperform (83 times) a traditional, disc-based
   DBMS. A more full-featured version of the system \cite{SADM08} that is able to
   execute across multiple machines within a local area cluster has
   been presented in August 2008. The data storage in H-store  is
   managed by a single-thread execution engine that resides
   underneath the transaction manager. Each individual site executes
   an autonomous instance of the storage engine with a fixed amount
   of memory allocated from its host machine. Multi-side nodes do
   not share any data structures with collocated sites, and hence
   there is no need to use concurrent data structures (every
   read-only table is replicated on all nodes nd other tables are
   divided horizontally into disjoint partitions  with a k-safety
   factor two). Thus, H-store  (at http://hstore.cs.brown.edu/documentation/architecture-overview/) was designed as a parallel, row-storage
   relational DBMS that runs on a cluster of shared-nothing, main memory
   executor nodes. The commercial version of H-store's design is
   VoltDB.\\
   More recently, during 2010 and 2011, Stonebraker has been a critic of the NoSQL movement
   \cite{Ston10,SADM10}: "Here, we argue that using MR systems to perform
tasks that are best suited for DBMSs yields less than satisfactory
results \cite{SADM10a}, concluding that MR is more like an
extract-transform-load (ETL) system than a DBMS, as it quickly loads
and processes large amounts of data in an ad hoc manner. As such, it
complements DBMS technology rather than competes with it." After a
number of arguments about MR (MapReduction) w.r.t. SQL (with GROUP
BY operation), the authors conclude that parallel DBMSs provide the
same computing model as MR (popularized by Google and Hadoop to
process key/value data pairs), with the added benefit of using a
declarative SQL language. Thus, parallel DBMSs offer great
scalability over the range of nodes that customers desire, where all
parallel DBMSs operate (pipelining) by creating a query plan that is
distributed to the appropriate nodes at execution time. When one
operator in this plan send data to next (running on the same or a
different node), the data are pushed by the first to the second
operator (this concept is analog to the process described in my book
\cite{Majk14}, February 2014, in Section 5.2.1 dedicated to
normalization of SQL terms (completeness of the
Action-relational-algebra category \textbf{RA}), so that
(differently from MR), the intermediate data is never written to
disk. The formal theoretical framework (the database category
\textbf{DB}) of the parallel DBMSs and the semantics of database
mappings between them is provided in Big Data integration theory as
well \cite{Majk14}.\\
It is interesting that in  \cite{SADM10}, the authors conclude that
parallel DBMSs excel at efficient querying of large data sets while
MR key/value style systems excel at complex analytics and ETL tasks,
and propose: "The result is a much more efficient overall system
than if one tries to do the entire application in either system.
That is, "smart software" is always a good idea."\\
The aim of this paper is to go one step in advance in developing
this NewSQL approach, and to extend the "classic" RDB systems with
both features:  to offer, on user's side, the standard RDB database
schema for SQL querying and, on computational side, the "vectorial"
relational database able to efficiently support the low-level
key/value data structures together, in the same logical SQL
framework. Moreover, this parsing of the standard RDBs into a
"vectorial" database efficiently resolves also the problems of NoSQL
applications with sparse-matrix
and "Open schema" data models.\\
The plan of this paper is the following: In Section 2 we present the
method of parsing of any RDB into a vector relation of the key/value
structure, compatible with most Big Data structures, and Open schema
solutions, but with preserving the RDB user-defined schema for the
software applications. We show that such a parsing changes the
standard semantics of the RDBs based on the FOL by introducing the
intensional concepts for user-defined relational tables.
Consequently, in Section 3 we introduce a \emph{conservative
intensional extension} of the FOL adequate to express the semantics
for the IRDBs and the SQL. In Section 4 we define a new semantics
for the IRDBSs and their canonical models based on the Data
Integration systems, where the user-defined RDB is a global schema
and the source schema is composed by the unique vector relations
which contains the parsed data of the whole used-defined RDB. As in
GAV (Global-As-View) Data Integration systems, we dematerialize the
global schema (i.e., user-defined RDB) and, in Section 5, we define
a \emph{query-rewriting algorithm} to translate the original query
written for the user-defined RDB schema into the source database
composed by the vector relation containing the parsed data.

%(M.I.T., Brown and Yale University)
%----------------------------------------------------------------------
% SECTION I: Introduction
%----------------------------------------------------------------------
\section{Vector databases with intensional FOL Semantics \label{sec:vector}}
In what follows, we denote by $B^A$ the set of all functions from
$A$ to $B$, and by $A^n$ a n-folded cartesian product $A \times
...\times A$ for $n \geq 1$,  we denote by $\neg, \wedge, \vee,
\Rightarrow$ and $\Leftrightarrow$ the logical operators negation,
conjunction, disjunction, implication and equivalence, respectively.
For any two logical formulae $\phi$ and $\psi$ we define the XOR
logical operator $\underline{\vee}$ by  $\phi \underline{\vee} \psi$
logically equivalent to $(\phi \vee \psi) \wedge \neg (\phi \wedge
\psi)$. Then we will use the following RDB definitions, based on the
standard First-Order Logic (FOL) semantics:
\begin{itemize}
  \item  A \emph{database schema} is a pair $\A = (S_A , \Sigma_A)$ where $S_A$ is
  a countable set of relational symbols (predicates in FOL) $r\in \mathbb{R}$
  with finite arity
   $n = ar(r) \geq 1$ ($~ar:\mathbb{R} \rightarrow \mathcal{N}$),
disjoint from a countable infinite set $\textbf{att}$ of attributes
(a domain of $a\in \textbf{att}$ is a
 nonempty finite subset $dom(a)$ of a countable set of individual symbols
$\textbf{dom}$). For any $r\in \mathbb{R}$, the sort of $r$, denoted
by tuple $\textbf{a} = atr(r)= <atr_r(1),...,atr_r(n)>$ where all
$a_i = atr_r(m) \in \textbf{att}, 1\leq m \leq n$, must be distinct:
if we use two equal domains for different attributes then we denote
them by $a_i(1),...,a_i(k)$ ($a_i$ equals to $a_i(0)$). Each index
("column") $i$, $1\leq i \leq ar(r)$, has a distinct column name
$nr_r(i) \in SN$ where $SN$ is the set of names with $nr(r) =
<nr_r(1),...,nr_r(n)>$. A relation symbol $r \in \mathbb{R}$
represents the \emph{relational name} and can be used as an atom
$r(\textbf{x})$ of FOL with variables in $\textbf{x}$ assigned to
its columns, so that $\Sigma_A$ denotes a set of sentences (FOL
formulae without free variables) called \emph{integrity constraints}
\index{integrity constraints} of the sorted FOL with sorts in
$\textbf{att}$.
%We define the function $nr:S_A \rightarrow SN$, such that for each relational symbol $r \in
%S_A$, $nr(r)$ is the unique name of this relation in the RDB schema$\A$.
%
\item An \emph{instance-database} of a nonempty schema  $\A$ is given by
$A = (\A,I_T) = \{R =\|r\| = I_T(r) ~|~r \in S_A \}$ where $I_T$ is
a Tarski's FOL interpretation which satisfies \emph{all} integrity
constraints in $\Sigma_A$ and maps a relational symbol $r \in S_A$
into an n-ary relation $R=\|r\|\in A$. Thus, an instance-database
$A$ is a set of n-ary relations, managed by  relational database
systems. \\ Let $A$ and $A' = (\A,I_T')$ be two instances of $\A$,
then a function $h:A \rightarrow A'$ is a \emph{homomorphism} from
$A$ into $A'$ if for every k-ary  relational symbol $r \in S_A$ and
every tuple $<v_1,...,v_k>$ of this k-ary relation in $A$,
$<h(v_1),...,h(v_k)>$ is a tuple of the same  symbol $r$ in $A'$. If
$A$ is an instance-database and $\phi$ is a sentence then we write
$A\models \phi~$ to mean that $A$ satisfies $\phi$. If $\Sigma$ is a
set of sentences then we write $A \models \Sigma$ to mean that
$A\models \phi$  for every sentence $\phi \in \Sigma$. Thus the set
of all instances of $\A$ is defined by $Inst(\A) = \{ A~|~ A \models
\Sigma_A \}$.
% We denote the set of all values in $A$ by $val(A)
%\subseteq \U$. Then 'atomic database' $\mathfrak{J}_A = \{\{<v_i>\}
%~|~ v_i \in val(A)\}$ is infinite iff $SK \subseteq val(A)$. Note
%that for each $a \in atr(r)$, a subset $dom(a) \subseteq
%\textbf{dom}$ is finite, and any introduction of Skolem constants is
%ordered $\omega_0,\omega_1,..$.
%
\item We consider a rule-based
\emph{conjunctive query} over a database schema $\A$ as an
expression $ q(\textbf{x})\longleftarrow r_1(\textbf{u}_1), ...,
r_n(\textbf{u}_n)$, with finite $n\geq 0$, $r_i$ are the relational
symbols (at least one) in $\A$ or the built-in predicates (e.g.
$\leq, =,$ etc.), $q$ is a relational symbol not in $\A$ and
$\textbf{u}_i$ are free tuples (i.e., one may use either variables
or constants). Recall that if $\textbf{v} = (v_1,..,v_m)$ then
$r(\textbf{v})$ is a shorthand for $r(v_1,..,v_m)$. Finally, each
variable occurring in $\textbf{x}$ is a \emph{distinguished}
variable that must also occur at least once in
$\textbf{u}_1,...,\textbf{u}_n$. Rule-based conjunctive queries
(called rules) are composed of a subexpression $r_1(\textbf{u}_1),
...., r_n(\textbf{u}_n)$ that is the \emph{body}, and
 the \emph{head} of this rule $q(\textbf{x})$ . The $Yes/No$
conjunctive queries are the rules with an empty head. If we can find
values for the variables of the rule, such that the body is
logically satisfied, then we can deduce the head-fact.
 This concept is captured by a notion of "valuation".\\
 The deduced head-facts  of a conjunctive query $q(\textbf{x})$ defined over an instance $A$ (for a given Tarski's
 interpretation $I_T$ of  schema $\A$) are
 equal to $\|q(x_1,...,x_k)\|_A = \{<v_1,...,v_k> \in \textbf{dom}^k ~|~  A \models \exists \textbf{y}(r_1(\textbf{u}_1)\wedge
 ...\wedge
r_n(\textbf{u}_n))[x_i/v_i]_{1\leq i \leq k} \}
 = I_T^*(\exists \textbf{y}(r_1(\textbf{u}_1)\wedge ...\wedge
r_n(\textbf{u}_n)))$, where the $\textbf{y}$ is a set of variables
which are not in the head of query, and $I_T^*$ is the unique
extension of $I_T$ to all formulae.
 We
recall that the conjunctive queries are monotonic and satisfiable,
and that a (Boolean) query is a class of instances that is closed
under isomorphism \cite{ChHa82}. Each conjunctive query corresponds
to a "select-project-join" term $t(\textbf{x})$
 of SPRJU algebra obtained from the formula $\exists \textbf{y}(r_1(\textbf{u}_1)\wedge ...\wedge
r_n(\textbf{u}_n))$, as explained in Section \ref{sec:newSQL}.
 \item
  We consider a finitary \emph{view} as a union of a finite set $S$ of conjunctive  queries with the same
 head $q(\textbf{x})$ over a schema $\A$, and from the equivalent
algebraic point of view, it is a "select-project-join-rename +
union" (SPJRU) finite-length term $t(\textbf{x})$ which corresponds
to union of the terms of conjunctive queries in $S$. In what follows
we will use the same notation for a FOL formula $q(\textbf{x})$ and
its equivalent algebraic SPJRU expression $t(\textbf{x})$. A
materialized view of an instance-database $A$ is an n-ary relation
$R = \bigcup_{q(\textbf{x}) \in S}\|q(\textbf{x})\|_A$. Notice that
a finitary view can  also have an infinite number of
    tuples. We denote the set of all  finitary materialized
views that can be obtained from an instance $A$ by $TA$.
 \end{itemize}
The principal idea is to use an analogy with a GAV Data Integration
\cite{Lenz02,Majk14} by using the database schema $\A = (S_A,
\Sigma_A)$ as a  global relational schema, used as a
user/application-program interface for  the query definitions in
SQL, and to represent the source database  of this Data Integration
system by  parsing of the RDB instance $A$ of the schema $\A$ into a
single vector relation $\overrightarrow{A}$. Thus, the original SQL
query $q(\textbf{x})$ has to be equivalently rewritten over
(materialized) source vector database $\overrightarrow{A}$.\\
The idea of a vector relation $\overrightarrow{A}$ for a given
relational database instance $A$ comes from the investigation of the
topological properties of the RDB systems, presented in Chapter 8 of
the book \cite{Majk14}. In order to analyze the algebraic lattice of
all RDB database instances each instance database $A$, composed by a
set of finitary relations $R_i \in A$, $i = 1,...,n$, in Lemma 21 is
defined the transformation of the instance database $A$ into a
vector relation $\overrightarrow{A}$, with $\overrightarrow{A} =
\bigcup_{R \in A} \overrightarrow{R}$ where for each relation $R$
$ar(R) \geq 1$  it the arity (the number of its columns) of this
relational table and $\pi_i$ is its $i$-th column projection, and
hence $\overrightarrow{R} = \bigcup_{1\leq i \leq ar(R)} \pi_i(R)$.
Such vectorial representation of a given database $A$ in
\cite{Majk14} is enough to define the lattice of RDB lattices,
because we do not needed the converse process (to define a database
$A$ from its vectorial representation).\\
However, by considering that a database $A$ is seen by the users and
their software applications (with the embedded SQL statements),
while $\overrightarrow{A}$ is its single-table internal
representation, over which is executed a  rewritten user's query,
the extracted information has to be converted in the RDB form w.r.t.
the relational schema of the original user's model. Consequently, we
need a reacher version of the vector database, such that we can
obtain an equivalent inverse transformation of it into the standard user defined RDB schema.\\
 In fact, each $i$-th column value  $d_i$ in a tuple $\textbf{d} = \langle d_1,...,d_i,...,d_{ar(r)} \rangle$
   of a relation $R_k = \|r_k\|, r_k \in S_A$,  of the
instance database $A$ is determined by the free dimensional
coordinates: relational name $nr(r)$, the attribute name $nr_r(i)$
of the i-th column, and the tuple index $Hash(\textbf{d})$ obtained
by hashing the string of the tuple $\textbf{d}$. Thus, the
relational schema of the vector relation is composed by the four
attributes, relational name, tuple-index, attribute name, and value,
i.e., \verb"r-name",   \verb"t-index", \verb"a-name" and
\verb"value", respectively, so that if we assume $r_V$ (the name of
the database $\A$) for the name of this vector relation
$\overrightarrow{A}$ then this
relation can be expressed by the quadruple\\
$r_V$(\verb"r-name",   \verb"t-index", \verb"a-name",
\verb"value"),\\
and the parsing of any RDB instance $A$ of a schema $\A$ can be
defined as:
\begin{definition} \textsc{Parsing RDB instances:}
\label{def:parsing} \\Given a database instance $A =
\{R_1,...,R_n\}$, $n\geq 1$, of a RDB schema $\A = (S_A,\Sigma_A)$
with $S_A = \{r_1,...,r_n\}$ such that $R_k = \|r_k\|, k = 1,...,n$,
then the extension $\overrightarrow{A} = \|r_V\|$ of the vector
relational symbol (name) $r_V$ with the schema $r_V$(\verb"r-name",
  \verb"t-index", \verb"a-name",
\verb"value"),  and \emph{NOT NULL} constraints for all its four
attributes,
and with the primary key composed by the first three attributes, is defined by:\\
we define the  operation \emph{\underline{PARSE}} for  a tuple
$\textbf{d} =\langle d_1,...,d_{ar(r_k)}\rangle$ of the relation
$r_k \in S_A$ by the
mapping\\
 $(r_k,\textbf{d}) ~~\mapsto ~~\{\langle r_k,
 Hash(\textbf{d}),nr_{r_k}(i),d_i \rangle |~  d_i \emph{NOT NULL},
1\leq i \leq ar(r_k)\}$,
so that\\
 (1) $~~~\overrightarrow{A} = \bigcup_{r_k \in S_A,\textbf{d} \in
 \|r_k\|} \emph{\underline{PARSE}}(r_k,\textbf{d})$.\\
 Based on the vector database representation
 $\|r_V\|$ we define a GAV Data Integration system  $\I = \langle \A, \S,
\M \rangle$ with the global schema $\A = (S_A, \Sigma_A)$, the
source schema $\S = (\{r_V\},\emptyset)$, and the set of mappings
$\M$ expressed by the tgds (tuple generating
dependencies)\\
(2) $~~~\forall y,x_1,...,x_{ar(r_k)}(((r_V(r_k,y,nr_{r_k}(1),x_1)
~\underline{\vee}~ x_1 \emph{NULL})\wedge ...\\...\wedge
(r_V(r_k,y,nr_{r_k}(ar(r_k)),x_{ar(r_k)})~\underline{\vee}~
x_{ar(r_k)} \emph{NULL})) \Rightarrow r_k(x_1,...,x_{ar(r_k)}))$,\\
for each $r_k \in S_A$.
% \\A model of  $\overrightarrow{\A}$ is any its
%instance, such that $\|r_V\|$ (considered as a Source database of
%this Data Integration system)   is equal to the parsing of a model
%$A$ of $\A$ and $\|r_k\| = \emptyset$ for all $r_k \in S_A$ (i.e.,
%all relations in the Global schema $\A$ are empty).
\end{definition}
The operation \underline{PARSE} corresponds to the parsing of the
tuple $\textbf{v}$ of the relation $r_k \in S_A$ of the user-defined
database schema $\A$ into a number of tuples of the vector relation
$r_V$. In fact, we can use this operation for virtual
inserting/deleting of the tuples in the user defined schema $\A$,
and  store them only in the vector relation $r_V$. This operation
avoids to materialize the user-defined (global) schema, but only the
source database $\S$, so that each user-defined SQL query has to be
equivalently rewritten over the source database (i.e., the big table
$\overrightarrow{A} = \|r_V\|$) as in standard
FOL Data Integration systems.\\
 Notice that
this parsing defines a kind of GAV Data Integration systems, where
the source database $\S$ is composed by the unique
%by-NULL-completed vector relation $\|\textbf{r}_V\|$.
vector relation
%It means that the vector relation
$\|r_V\| = \overrightarrow{A}$ (Big Data) which does not contain
NULL values, so that we do not unnecessarily save the NULL values of
the user-defined relational tables $r_k \in S_A$ in the main
memories of the parallel RDBMS used to horizontal partitioning of
the unique big-table $\overrightarrow{A}$. Moreover, any adding of
the new columns to the user-defined schema $\A$ does not change the
table $\overrightarrow{A}$, while the deleting of a $i$-th column of
a relation $r$ will delete all tuples $r_V(x,y,z,v)$ where $x =
nr(r)$ and $z = nr_r(i)$ in the main memory of the parallel RDBMS.
Thus, we obtain very \emph{schema-flexible} RDB model for Big
Data.\\
Other obtained NoSQL systems properties are:
\begin{itemize}
  \item \emph{Compatible with key/value systems}.Note that the vector big-table
$\overrightarrow{A}$ is in the 6th normal form, that is with the
primary key corresponding to the first three attributes (the free
dimensional coordinates) and the unique value attribute. Thus we
obtained the key/value style used for NoSQL Big Data systems. That
is, the RDB parsing with resulting Data Integration system subsumes
all Big Data key/value systems.\\
  \item \emph{Compatible with "Open schema" systems}. Entity-attribute-value model (EAV)
   is a data model to describe entities where the number of attributes
   (properties, parameters) that can be used to describe them is potentially vast,
   but the number that will actually apply to a given entity is relatively
   modest. In mathematics, this model is known as a sparse matrix.
   EAV is also known as object-attribute-value model, vertical database model and open schema.
   We can use the special relational symbol with name "\verb"OpenSchema""
   in the user database schema $\A$ so that its tuples in $\overrightarrow{A}$
   will corresponds to atoms $r_V(\verb"OpenSchema", object,
   attribute, value)$. In this case the software
   developed for the applications which use the Open schema data
   will directly access to the vector relation $\overrightarrow{A}$
   and DBMS will restrict all operations only to tuples where the
   first attribute has the value equal to \verb"OpenSchema" (during an inserting
   of  a new tuple $\langle object, attribute, value \rangle$ the DBMS
   inserts also the value \verb"OpenSchema" in the first column of
   $\overrightarrow{A}$).
  \end{itemize}
But this simple and unifying framework needs more investigation for
the SQL and underlying logical framework. In fact, we can easy see
that the mapping tgds used from the Big Data vector table
$\overrightarrow{A}$ (the source schema in Data Integration) into
user-defined RDB schema $\A$ (the global schema of this Data
Integration system with integrity constraints) is not simple FOL
formula. Because the same element $r_k$ is used as a predicate
symbol (on the right-side of the tgd's implication) and as a value
(on the left side of the implication as the first value in the
predicate $r_V$). It means that the elements of the domain of this
logic are the elements of other classes and are the classes for
themselves as well. Such semantics is not possible in the standard
FOL, but only in the \emph{intensional} FOL, and hence the Data
Integration $\I$ is not a classic FOL Data Integration as in
\cite{Lenz02} but an Intensional Data Integration system. In the
next sections we will investigate what is the proper logic framework
for this class of RDBs, denominated as IRDBs (Intensional RDBs), and
to show that the standard SQL is complete in this new logical
framework.

\section{Conservative intensional extension of the FOL for IRDBs}
 The first conception of intensional entities (or concepts) is built
into the \emph{possible-worlds} treatment of Properties, Relations
and Propositions (PRP)s. This conception is commonly attributed to
Leibniz, and underlies Alonzo Church's alternative formulation of
Frege's theory of senses ("A formulation of the Logic of Sense and
Denotation" in Henle, Kallen, and Langer, 3-24, and "Outline of a
Revised Formulation of the Logic of Sense and Denotation" in two
parts, Nous,VII (1973), 24-33, and VIII,(1974),135-156). This
conception of PRPs is ideally suited for treating the
\emph{modalities} (necessity, possibility, etc..) and to Montague's
definition of intension of a given virtual predicate
$\phi(x_1,...,x_k)$ (a FOL open-sentence with the tuple of free
variables $(x_1,...x_k)$), as a mapping from possible worlds into
extensions of this virtual predicate. Among the possible worlds we
distinguish the \emph{actual} possible world. For example, if we
consider a set of predicates, of a given Database,
and their extensions in different time-instances, then the actual possible world is identified by the current instance of the time.\\
The second conception of intensional entities is to be found in
Russell's doctrine of logical atomism. In this doctrine it is
required that all complete definitions of intensional entities be
finite as well as unique and non-circular: it offers an
\emph{algebraic} way for definition of complex intensional entities
from simple (atomic) entities (i.e., algebra of concepts),
conception also evident in Leibniz's remarks. In a predicate logics,
predicates and open-sentences (with free variables) expresses
classes (properties and relations), and sentences express
propositions. Note that classes (intensional entities) are
\emph{reified}, i.e., they belong to the same domain as individual
objects (particulars). This endows the intensional logics with a
great deal of uniformity, making it possible to manipulate classes
and individual objects in the same language. In particular, when
viewed as an individual
object, a class can be a member of another class.\\
 The distinction between
intensions and extensions is important (as in lexicography
\cite{PuBo93}), considering that extensions can be notoriously
difficult to handle in an efficient manner. The extensional equality
theory of predicates and functions under higher-order semantics (for
example, for two predicates with the same set of attributes $p = q$
is true iff these symbols are interpreted by the same relation),
that is, the strong equational theory of intensions, is not
decidable, in general. For example,  the second-order predicate
calculus and Church's simple theory of types, both under the
standard semantics, are not even semi-decidable. Thus, separating
intensions from extensions makes it possible to have an equational
theory over predicate and function names (intensions) that is
separate from the extensional equality of
relations and functions. \\
Relevant recent work  about the intension, and its relationship with
FOL, has been presented in \cite{Fitt04} in the consideration of
rigid and \emph{non-rigid} objects, w.r.t. the possible worlds,
where the rigid objects, like "George Washington", and are the same
things from possible world to possible world. Non-rigid objects,
like "the Secretary-General of United Nations", are varying from
circumstance to circumstance and can be modeled semantically by
functions from possible worlds to domain of rigid objects, like
intensional entities. However, Fitting substantially and ad-hock
changes the syntax and semantics of FOL, and introduces the
Higher-order Modal logics, differently from our approach. More about
other relevant recent works are presented in
\cite{Majk09FOL,Majk12a} where a new conservative intensional
extension of the
Tarski's semantics of the FOL is defined.\\
 %
%\section{Intensional FOL language with intensional abstraction} \label{section:intensional}
%\label{section:intensional}
%
 Intensional entities are such concepts as
propositions and properties. The term 'intensional' means that they
violate the principle of extensionality; the principle that
extensional equivalence implies identity. All (or most) of these
intensional entities have been classified at one time or another as
kinds of Universals \cite{Beal93}.\\
We consider a non empty domain $~\D = D_{-1} \bigcup D_I$,  where a
subdomain $D_{-1}$ is made of
 particulars (extensional entities), and the rest $D_I = D_0 \bigcup
 D_1 ...\bigcup D_n ...$ is made of
 universals ($D_0$ for propositions (the 0-ary concepts), and  $D_n, n \geq 1,$ for
 n-ary concepts).\\
 The fundamental entities
are \emph{intensional abstracts} or so called, 'that'-clauses. We
assume that they are singular terms; Intensional expressions like
'believe', mean', 'assert', 'know',
 are standard two-place predicates  that take 'that'-clauses as
 arguments. Expressions like 'is necessary', 'is true', and 'is
 possible' are one-place predicates that take 'that'-clauses as
 arguments. For example, in the intensional sentence "it is
 necessary that $\phi$", where $\phi$ is a proposition, the 'that $\phi$' is
 denoted by the $\lessdot \phi \gtrdot$, where $\lessdot \gtrdot$ is the intensional abstraction
 operator which transforms a logic formula into a \emph{term}. Or, for example, "x believes that $\phi$" is given by formula
$p_i(x,\lessdot \phi \gtrdot)$ ( $p_i$ is binary 'believe'
predicate). We introduce an  intensional FOL  \cite{Majk12a}, with
slightly different intensional abstraction than that originally
presented  in \cite{Beal79}, as follows:
 %$\vspace*{-3mm}$
 \begin{definition} \label{def:bealer}
  The syntax of the First-order Logic (FOL) language $\L$ with intensional abstraction
$\lessdot \gtrdot$ is as follows:\\
 Logical operators $(\wedge, \neg, \exists)$; Predicate letters $r_i,p_i \in
 \mathbb{R}$
 with a given arity $k_i = ar(r_i) \geq 1$, $i = 1,2,...$ (the functional letters are considered as particular case of the predicate
 letters); a set PR  of propositional letters (nullary predicates) with a truth $r_\emptyset \in PR \bigcap \mathbb{R}$;  Language constants $\overline{0}, \overline{1},...,\overline{c},\overline{d}...$; Variables $x,y,z,..$ in $\V$; Abstraction $\lessdot \_ \gtrdot$, and punctuation
 symbols (comma, parenthesis).
 With the following simultaneous inductive definition of \emph{term} and
 \emph{formula}:\\
 %\begin{enumerate}
   1. All variables and constants  are terms. All propositional letters are formulae.\\
   2. If $~t_1,...,t_k$ are terms then $r_i(t_1,...,t_k)$ is a formula
 for a k-ary predicate letter $r_i \in \mathbb{R}$ .\\
   3. If $\phi$ and $\psi$ are formulae, then $(\phi \wedge \psi)$, $\neg \phi$, and
 $(\exists x)\phi$ are formulae. \\
   4. If $\phi(\textbf{x})$ is a formula (virtual predicate) with a list of free variables in $\textbf{x} =(x_1,...,x_n)$ (with ordering
from-left-to-right of their appearance in $\phi$), and  $\alpha$ is
its sublist of \emph{distinct} variables,
 then $\lessdot \phi \gtrdot_{\alpha}^{\beta}$ is a term, where $\beta$ is the remaining list of free variables preserving ordering in $\textbf{x}$ as well. The externally quantifiable variables are the \emph{free} variables not in $\alpha$.
  When $n =0,~ \lessdot \phi \gtrdot$ is a term which denotes a
proposition, for $n \geq 1$ it denotes
 a n-ary concept.\\
% \end{enumerate}
An occurrence of a variable $x_i$ in a formula (or a term) is
\emph{bound} (\emph{free}) iff it lies (does not lie) within a
formula of the form $(\exists x_i)\phi$ (or a term of the form
$\lessdot \phi \gtrdot_{\alpha}^{\beta}$ with $x_i \in \alpha$). A
variable is free (bound) in a formula (or term) iff it has (does not
have) a free occurrence in that formula (or term). A \emph{sentence}
is a formula having no free variables.
%$\vspace*{-3mm}$
\end{definition}
An interpretation (Tarski) \index{Tarski's interpretations} $I_T$
consists of a nonempty domain
   $\D = D_{-1} \bigcup D_I$ and a mapping that assigns to any predicate letter $r_i \in
   \mathbb{R}$ with $k = ar(r_i)\geq 1$, a relation $\|r_i\| = I_T(r_i) \subseteq
   \D^k$;
   to each
   individual constant $\overline{c}$ one given element $I_T(\overline{c}) \in
   \D$, with $I_T(\overline{0}) = 0, I_T(\overline{1}) = 1$ for
   natural numbers $\N =\{0,1,2,...\}$, and to any
   propositional letter $p \in PR$ one  truth value $I_T(p) \in
    \{f,t\}$, where $f$ and $t$ are the empty set $\{\}$
   and the singleton set $\{<>\}$ (with the empty tuple $<> \in
   D_{-1}$), as those used  in the Codd's relational-database algebra \cite{Codd72} respectively,
    so that for any $I_T$, $I_T(r_\emptyset) = \{<>\}$
   (i.e., $r_\emptyset$ is a tautology), while $Truth \in D_0$ denotes the concept (intension)
of this tautology. \\
%--------------------------------------
Note that in the intensional semantics a k-ary functional symbol,
for $k \geq 1$, in standard (extensional) FOL is considered as a
$(k+1)$-ary predicate symbols: let $f_m$ be such a $(k+1)$-ary
predicate symbol which represents a k-ary function denoted by
$\underline{f}_m$ with standard Tarski's interpretation
$I_T(\underline{f}_m):\D^k \rightarrow \D$. Then $I_T(f_m)$ is a
relation obtained from its graph, i.e.,  $I_T(f_m) = R =
\{(d_1,...,d_k,I_T(\underline{f}_m)(d_1,...,d_k)) ~| ~d_i \in \D,
1\leq i \leq k \}$.
%----------------------------------------
The universal quantifier is defined by $\forall = \neg \exists
\neg$. Disjunction $\phi \vee \psi$ and implication $\phi
\Rightarrow \psi$ are expressed by
 $\neg(\neg \phi \wedge \neg \psi)$ and $\neg \phi \vee
 \psi$, respectively.
 In FOL with the
identity $\doteq$, the formula $(\exists_1 x)\phi(x)$ denotes the
formula $(\exists x)\phi(x) \wedge (\forall x)(\forall y)(\phi(x)
\wedge \phi(y)  \Rightarrow (x \doteq y))$. We denote by $R_{=}$ the
Tarski's interpretation of $\doteq$. In what follows any
open-sentence, a formula $\phi$ with non empty tuple  of free
variables $(x_1,...,x_m)$, will be called a m-ary
  \emph{virtual predicate}, denoted also by
$\phi(x_1,...,x_m)$. This definition contains the precise method of
establishing the \emph{ordering} of variables in this tuple:
  such an method that will be adopted here is the ordering of appearance, from left to right, of free variables in $\phi$.
   This method of composing the tuple of free variables
  is the unique and canonical way of definition of the virtual predicate from a given
  formula.\\
An \emph{intensional interpretation} of this intensional FOL is a
mapping between the set $\L$ of formulae of the logic language  and
 intensional entities in $\D$, $I:\L \rightarrow \D$, is a kind of
 "conceptualization", such that  an open-sentence (virtual
 predicate)
 $\phi(x_1,...,x_k)$ with a tuple $\textbf{x}$ of all free variables
 $(x_1,...,x_k)$ is mapped into a k-ary \emph{concept}, that is, an intensional entity  $u =
 I(\phi(x_1,...,x_k)) \in D_k$, and (closed) sentence $\psi$ into a proposition (i.e., \emph{logic} concept) $v =
 I(\psi) \in D_0$ with $I(\top) = Truth \in D_0$ for a FOL tautology $\top$.
This interpretation $I$ is extended also to the terms (called as
denotation as well).
 A language constant $\overline{c}$ is mapped into a
 particular (an extensional entity) $a = I(\overline{c}) \in D_{-1}$ if it is a proper name, otherwise in a correspondent concept in
$\D$.
%------------new-----------------
%A variable $x_i$ is mapped into the attribute-name $I(x_i) \in
%D_{-1}$ of this variable, while for each $k$-ary  atom
%$r_i(\textbf{x})$, $I(\lessdot r_i(\textbf{x})
%\gtrdot_{\textbf{x}})$ is the relation-name (symbol) $r_i \in
%\mathbb{R}$ (only if $r_i$ is not  defined as a language constant as well).
For each $k$-ary  atom $r_i(\textbf{x})$, $I(\lessdot
r_i(\textbf{x}) \gtrdot_{\textbf{x}})$ is the relation-name (symbol)
$r_i \in \mathbb{R}$ (only if $r_i$ is not  defined as a language
constant as well).
%------------------------------------------
The extension of $I$ to  the complex abstracted terms  is given in \cite{Majk12a} (in Definition 4).\\
 An assignment $g:\V \rightarrow \D$ for variables in $\V$ is
applied only to free variables in terms and formulae.  Such an
assignment $g \in \D^{\V}$ can be recursively uniquely extended into
the assignment $g^*:\T X \rightarrow \D$, where $\T X$ denotes the
set of all terms with variables in $X \subseteq \V$ (here $I$ is an
intensional interpretation of this FOL, as explained
in what follows), by :\\
1. $g^*(t_k) = g(x) \in \D$ if the term $t_k$ is a variable $x \in
\V$.\\
2. $g^*(t_k) = I(\overline{c}) \in \D$ if the term $t_k$ is a
constant
$\overline{c}$.\\
3. if $t_k$ is an abstracted term $\lessdot \phi
\gtrdot_{\alpha}^{\beta}$,  then $g^*(\lessdot \phi
\gtrdot_{\alpha}^{\beta}) = I(\phi[\beta /g(\beta)] ) \in D_k, k =
|\alpha|$ (i.e., the number of variables in $\alpha$), where
$g(\beta) = g(y_1,..,y_m) = (g(y_1),...,g(y_m))$ and $[\beta
/g(\beta)]$ is a uniform replacement of each i-th variable in the
list $\beta$
with the i-th constant in the list $g(\beta)$. Notice that $\alpha$ is the list of all free variables in the formula $\phi[\beta /g(\beta)]$.\\
  We denote by $~t_k/g~$ (or $\phi/g$) the ground term (or
formula) without free variables, obtained by assignment $g$ from a
term $t_k$ (or a formula $\phi$), and by  $\phi[x/t_k]$ the formula
obtained by  uniformly replacing $x$ by a term $t_k$ in $\phi$.\\
The distinction between intensions and extensions is important
 especially because we are now able to have an \emph{equational
 theory} over intensional entities (as  $\lessdot \phi \gtrdot$), that
 is predicate and function "names", that is separate from the
 extensional equality of relations and functions.
% \end{center}
 An \emph{extensionalization function} $h$ assigns to the intensional elements of $\D$ an appropriate
extension as follows: for each proposition $u \in D_0$, $h(u) \in
 \{f,t\} \subseteq \P(D_{-1})$ is its
 extension (true or false value); for each n-ary
 concept $u \in D_n$, $h(u)$ is a subset of $\D^n$
 (n-th Cartesian product of $\D$); in the case of particulars $u \in
 D_{-1}$, $h(u) = u$.\\
 We define  $\D^0 = \{<>\}$, so that $\{f,t\} = \P(\D^0)$, where $\P$ is the powerset operator.
 Thus we have (we denote the disjoint union by '+'):
%\begin{center}
 \\$h = (h_{-1}  + \sum_{i\geq 0}h_i):\sum_{i
\geq -1}D_i \longrightarrow D_{-1} +  \sum_{i\geq 0}\P(D^i)$,\\
%\end{center}
 where $h_{-1} = id:D_{-1} \rightarrow D_{-1}$
is identity mapping, the mapping $h_0:D_0 \rightarrow \{f,t\}$
assigns the truth values in $ \{f,t\}$ to all propositions, and the
mappings $h_i:D_i \rightarrow \P(D^i)$, $i\geq 1$, assign an
extension to all concepts. Thus, the intensions can be seen as
\emph{names} of abstract or concrete entities, while the extensions
correspond to
various rules that these entities play in different worlds.\\
\textbf{Remark:} (Tarski's constraints) This intensional semantics
has to preserve standard Tarski's semantics of the FOL. That is, for
any formula $\phi \in \L$ with a tuple of free variables
$(x_1,...,x_k)$,  and  $h \in \E$,  the following conservative
 conditions for all  assignments $g,g' \in \D^{\V}$ has to be satisfied: \\
(T)$~~~~~~~h(I(\phi/g)) = t~~$ iff
$~~(g(x_1),...,g(x_k)) \in h(I(\phi))$;\\
and, if $\phi$ is a predicate letter $p$, $k = ar(p) \geq 2$ which
represents a (k-1)-ary
functional symbol $f^{k-1}$ in standard FOL,\\
 (TF)$~~~~~~~h(I(\phi/g)) = h(I(\phi/g')) = t$ and
 $\forall_{1\leq i \leq k-1}(g'(x_i) = g(x_i))~~$ implies $~~g'(x_{k+1})=
 g(x_{k+1})$.\\
$\square$\\
%We define the following binary relation:\\  $R_{=} =
%\{(g(x),g(y))~|~ g \in \D^{\V}$ and $h(I(p_1^2(x,y)/g)) = t \})$
 Thus, intensional  FOL  has a simple Tarski's
first-order semantics, with a decidable
 unification problem, but we need also the actual world mapping
 which maps any intensional entity to its \emph{actual world
 extension}. In what follows we will identify a \emph{possible world} by a
 particular mapping which assigns, in such a possible world,  the extensions to intensional entities.
 This is direct bridge between
 an intensional FOL  and a possible worlds representation
 \cite{Lewi86,Stal84,Mont70,Mont73,Mont74,Majk09FOL}, where the intension (meaning) of a proposition is a
 \emph{function}, from a set of possible worlds $\W$ into the set of
 truth-values.
 % and  properties and functions from $\W$ to sets of possible (usually not-actual) objects.\\
 Consequently, $\E$ denotes the set of possible
\emph{extensionalization functions} $h$ satisfying the constraint
(T). Each $h \in \E$ may be seen as a \emph{possible world}
(analogously to Montague's intensional semantics for natural
language \cite{Mont70,Mont74}), as it has been demonstrated in
\cite{Majk08in,Majk08ird}, and given by the bijection
%\begin{center}
$~~~is:\W \simeq \E$.\\
Now we are able to formally define this intensional semantics
\cite{Majk09FOL}:
 \begin{definition} \label{def:intensemant} \textsc{Two-step \textsc{I}ntensional
 \textsc{S}emantics:}\\
Let $~\mathfrak{R} = \bigcup_{k \in \mathbb{N}} \P(\D^k) =
\sum_{k\in \mathbb{N}}\P(D^k)$ be the set of all k-ary relations,
where $k \in \mathbb{N} = \{0,1,2,...\}$. Notice that $\{f,t\} =
\P(\D^0) \in \mathfrak{R}$, that is, the truth values are extensions
in $\mathfrak{R}$. The intensional semantics of the logic language
with the set of formulae $\L$ can be represented by the  mapping
\begin{center}
$~~~ \L ~\rTo^{I} \D ~\Longrightarrow_{w \in \W}~ \mathfrak{R}$,
\end{center}
where $\rTo^{I}$ is a \emph{fixed intensional} interpretation $I:\L
\rightarrow \D$ and $~\Longrightarrow_{w \in \W}~$ is \emph{the set}
of all extensionalization functions $h = is(w):\D \rightarrow
\mathfrak{R}$ in $\E$, where $is:\W \rightarrow \E$ is the mapping
from the set of possible worlds to the set of
 extensionalization functions.\\
 We define the mapping $I_n:\L_{op} \rightarrow
\mathfrak{R}^{\W}$, where $\L_{op}$ is the subset of formulae with
free variables (virtual predicates), such that for any virtual
predicate $\phi(x_1,...,x_k) \in \L_{op}$ the mapping
$I_n(\phi(x_1,...,x_k)):\W \rightarrow \mathfrak{R}$ is the
Montague's meaning (i.e., \emph{intension}) of this virtual
predicate \cite{Lewi86,Stal84,Mont70,Mont73,Mont74}, that is, the
mapping which returns with the extension of this (virtual) predicate
in each possible world $w\in \W$.
\end{definition}
 Another relevant question w.r.t. this two-step
interpretations of an intensional semantics is how in it is managed
the extensional identity relation $\doteq$ (binary predicate of the
identity) of the FOL. Here this extensional identity relation is
mapped into the binary concept $Id = I(\doteq(x,y)) \in D_2$, such
that $(\forall w \in \W)(is(w)(Id) = R_{=})$, where $\doteq(x,y)$
(i.e., $p_1^2(x,y)$) denotes an atom of the FOL of the binary
predicate for identity in FOL, usually written by FOL formula $x
\doteq y$.\\
 Note that here we prefer to distinguish this \emph{formal
symbol} $~ \doteq ~ \in \mathbb{R}$ of the built-in identity binary
predicate letter in the FOL, from the standard mathematical
symbol '$=$' used in all mathematical definitions in this paper.\\
 In what follows we will use the function $f_{<>}:\mathfrak{R}
\rightarrow \mathfrak{R}$, such that for any relation $R \in
\mathfrak{R}$, $f_{<>}(R) = \{<>\}$ if $R \neq \emptyset$;
$\emptyset$ otherwise. Let us define the following set of algebraic
operators for
 relations in $\mathfrak{R}$:
\begin{enumerate}
\item binary operator $~\bowtie_{S}:\mathfrak{R} \times \mathfrak{R} \rightarrow
\mathfrak{R}$,
 such that for any two relations $R_1, R_2 \in
 \mathfrak{R}~$, the
 $~R_1 \bowtie_{S} R_2$ is equal
to the relation obtained by natural join
 of these two relations $~$ \verb"if"
 $S$ is a non empty
set of pairs of joined columns of respective relations (where the
first argument is the column index of the relation $R_1$ while the
second argument is the column index of the joined column of the
relation $R_2$); \verb"otherwise" it is equal to the cartesian
product $R_1\times R_2$.\\ For example, the logic formula
$\phi(x_i,x_j,x_k,x_l,x_m) \wedge \psi (x_l,y_i,x_j,y_j)$ will be
traduced by the algebraic expression $~R_1 \bowtie_{S}R_2$ where
$R_1 \in \P(\D^5), R_2\in \P(\D^4)$ are the extensions for a given
Tarski's interpretation  of the virtual predicate $\phi, \psi$
relatively, so that $S = \{(4,1),(2,3)\}$ and the resulting relation
will have the following ordering of attributes:
$(x_i,x_j,x_k,x_l,x_m,y_i,y_j)$.
\item unary operator $~ \sim:\mathfrak{R} \rightarrow \mathfrak{R}$, such that for any k-ary (with $k \geq 0$)
relation $R \in  \P(\D^{k}) \subset \mathfrak{R}$
 we have that $~ \sim(R) = \D^k \backslash R \in \D^{k}$, where '$\backslash$' is the substraction of relations. For example, the
logic formula $\neg \phi(x_i,x_j,x_k,x_l,x_m)$ will be traduced by
the algebraic expression $~\D^5 \backslash R$ where $R$ is the
extensions for a given Tarski's interpretation  of the virtual
predicate $\phi$.
\item unary operator $~ \pi_{-m}:\mathfrak{R} \rightarrow \mathfrak{R}$, such that for any k-ary (with $k \geq 0$) relation $R \in \P(\D^{k}) \subset \mathfrak{R}$
we have that $~ \pi_{-m} (R)$ is equal to the relation obtained by
elimination of the m-th column of the relation $R~$ \verb"if" $1\leq
m \leq k$ and $k \geq 2$; equal to $~f_{<>}(R)~$ \verb"if" $m = k
=1$; \verb"otherwise" it is equal to $R$. \\For example, the logic
formula $(\exists x_k) \phi(x_i,x_j,x_k,x_l,x_m)$ will be traduced
by the algebraic expression $~\pi_{-3}(R)$ where $R$ is the
extensions for a given Tarski's interpretation  of the virtual
predicate $\phi$ and the resulting relation will have the following
ordering of attributes: $(x_i,x_j,x_l,x_m)$.
\end{enumerate}
Notice that the ordering of attributes of resulting relations
corresponds to the method used for generating the ordering of
variables in the tuples of free variables adopted for virtual
predicates.
\begin{definition}  \label{def:int-algebra} Intensional algebra for the intensional FOL  in Definition \ref{def:bealer} is a structure $~\A_{int}
= ~(\D,  f, t, Id, Truth,  \{conj_{S}\}_{ S \in \P(\mathbb{N}^2)},
neg, \{exists_{n}\}_{n \in \mathbb{N}})$,  $~~$  with
%which contains  a non empty domain $~\D $;\\
 binary operations  $~~conj_{S}:D_I\times D_I \rightarrow D_I$,
   unary operation  $~~neg:D_I\rightarrow D_I$,  unary
   operations $~~exists_{n}:D_{I}\rightarrow D_I$,  such that for any
extensionalization function $h \in \E$,
and $u \in D_k, v \in D_j$, $k,j \geq 0$,\\
1. $~h(Id) = R_=~$ and $~h(Truth) = \{<>\}$.\\
2. $~h(conj_{S}(u, v)) = h(u) \bowtie_{S}h(v)$, where $\bowtie_{S}$
is the natural join operation defined above and $conj_{S}(u, v) \in
D_m$ where $m = k + j - |S|$
 if for every pair $(i_1,i_2) \in S$ it holds that $1\leq i_1 \leq k$, $1 \leq i_2 \leq j$ (otherwise $conj_{S}(u, v) \in D_{k+j}$).\\
3. $~h(neg(u)) = ~\sim(h(u)) = \D^k \backslash (h(u))$,
 where  $~\sim~$ is the operation
defined above and $neg(u) \in D_k$.\\
 4. $~h(exists_{n}(u)) =
\pi_{-n}(h(u))$, where $\pi_{-n}$ is the operation defined above and
\\ $exists_n(u) \in D_{k-1}$ if $1 \leq n \leq k$ (otherwise
$exists_n$ is the identity function).
\end{definition}
Notice that for $u \in D_0$, $~h(neg(u)) = ~\sim(h(u)) = \D^0
\backslash (h(u)) = \{<>\} \backslash (h(u)) \in \{f,t\}$.\\
Intensional interpretation $I:\L \rightarrow \D$ satisfies the
following homomorphic extension:
\begin{enumerate}
  \item The logic formula $\phi(x_i,x_j,x_k,x_l,x_m) \wedge \psi
(x_l,y_i,x_j,y_j)$ will be intensionally interpreted by the concept
$u_1 \in D_7$, obtained by the algebraic expression $~
conj_{S}(u,v)$ where $u = I(\phi(x_i,x_j,x_k,x_l,x_m)) \in D_5, v =
I(\psi (x_l,y_i,x_j,y_j))\in D_4$ are the concepts of the virtual
predicates $\phi, \psi$, relatively, and $S = \{(4,1),(2,3)\}$.
Consequently, we have that for any two formulae $\phi,\psi \in \L$
and a particular  operator $conj_S$ uniquely determined by tuples of
free variables in these two formulae, $I(\phi \wedge \psi) =
conj_{S}(I(\phi),I(\psi))$.
  \item The logic formula $\neg \phi(x_i,x_j,x_k,x_l,x_m)$ will be
intensionally interpreted by the concept $u_1  \in D_5$, obtained by
the algebraic expression $~neg(u)$ where $u =
I(\phi(x_i,\\x_j,x_k,x_l,x_m)) \in D_5$ is the concept of the
virtual predicate $\phi$. Consequently, we have that for any formula
$\phi \in \L$, $~I(\neg \phi) = neg(I(\phi))$.
  \item The logic formula $(\exists x_k) \phi(x_i,x_j,x_k,x_l,x_m)$ will
be intensionally interpreted by the concept $u_1  \in D_4$, obtained
by the algebraic expression $~exists_{3}(u)$ where $u =
I(\phi(x_i,x_j,x_k,x_l,x_m)) \in D_5$ is the concept of the virtual
predicate $\phi$. Consequently, we have that for any formula $\phi
\in \L$ and a particular operator $exists_{n}$ uniquely determined
by the position of the  existentially quantified variable in the
tuple of free variables in $\phi$ (otherwise $n =0$ if this
quantified variable is not a free variable in $\phi$), $~I((\exists
x)\phi) = exists_{n}(I(\phi))$.
\end{enumerate}
 Once one has found a method for specifying the interpretations of
singular terms of $\L$ (take in consideration the particularity of
abstracted terms), the Tarski-style definitions of truth and
validity for  $\L$ may be given in the customary way.
% An \emph{intensional interpretation} $I$ \cite{Beal82}
% maps each k-ary predicate letter  of
%$\L$ to k-ary concept in $D_k$. It can be extended to all formulae in usual way.
%What is being south specifically
What is proposed  specifically in \cite{Majk12a} is a method for
characterizing the intensional interpretations of singular terms of
$\L$ in such a way that a given singular abstracted term $\lessdot
\phi \gtrdot_{\alpha}^{\beta}$ will denote an appropriate property,
relation, or proposition, depending on the value of $m =
|\alpha|$.\\
 Notice than if $\beta = \emptyset$ is the empty
list, then $I(\lessdot \phi \gtrdot_{\alpha}^{\beta} ) = I(\phi)$.
Consequently, the denotation of $\lessdot \phi\gtrdot $
 is equal to the meaning of a proposition $\phi$, that is, $~I(\lessdot \phi\gtrdot) =
I(\phi)\in D_0$.  In the case when $\phi$ is an atom
$p_i(x_1,..,x_m)$ then $I (\lessdot
p_i(x_1,..,x_m)\gtrdot_{x_1,..,x_m}) = I(p_i(x_1,..,x_m)) \in D_m$,
while \\$I (\lessdot p_i(x_1,..,x_m)\gtrdot^{x_1,..,x_m}) = union
(\{I(p_i(g(x_1),..,g(x_m)))~|~ g \in \D^{\{x_1,..,x_m\}} \}) \in
D_0$,  with $h(I (\lessdot p_i(x_1,..,x_m)\gtrdot^{x_1,..,x_m})) =
h(I((\exists x_1)...(\exists x_m)p_i(x_1,..,x_m))) \in \{f,t\}$.\\
For example,\\ $h(I(\lessdot p_i(x_1) \wedge \neg p_i(x_1)
\gtrdot^{x_1})) = h(I((\exists x_1)(\lessdot p_i(x_1) \wedge \neg
p_i(x_1)
\gtrdot^{x_1}))) = f$.\\
The interpretation of a more complex abstract $\lessdot \phi
\gtrdot_\alpha^{\beta}$ is defined in terms of the interpretations
of the relevant syntactically simpler expressions, because the
interpretation of more complex formulae is defined in terms of the
interpretation of the relevant syntactically simpler formulae, based
on the intensional algebra above. For example, $I(p_i(x) \wedge
p_k(x)) = conj_{\{(1,1)\}}(I(p_i(x)), I(p_k(x)))$, $I(\neg
\phi) = neg(I(\phi))$, $I(\exists x_i)\phi(x_i,x_j,x_i,x_k) = exists_3(I(\phi))$.\\
Consequently, based on the intensional algebra in Definition
\ref{def:int-algebra} and on intensional interpretations of
abstracted terms, it holds that the interpretation of any formula in
$\L$ (and any abstracted term) will be reduced to an algebraic
expression over interpretations of primitive atoms in $\L$. This
obtained expression is finite for any finite formula (or abstracted
term), and represents the \emph{
meaning} of such finite formula (or abstracted term).\\
Let $\A_{FOL} = (\L, \doteq, \top, \wedge, \neg, \exists)$ be a free
syntax algebra for "First-order logic with identity $\doteq$", with
the set $\L$ of first-order logic formulae,  with $\top$ denoting
the tautology formula (the contradiction formula is denoted by $
\neg \top$), with the set of variables in $\V$ and the domain of
values in $\D$ . \\
Let us define the extensional relational algebra  for the FOL by,\\
$\A_{\mathfrak{R}} = (\mathfrak{R}, R_=, \{<>\}, \{\bowtie_{S}\}_{ S
\in \P(\mathbb{N}^2)}, \sim, \{\pi_{-n}\}_{n \in \mathbb{N}})$,
\\where $ \{<>\} \in \mathfrak{R}$ is the algebraic value
correspondent to the logic truth, and $R_=$ is the binary relation
for extensionally equal elements.
We  use '$=$' for the extensional identity for relations in $\mathfrak{R}$.\\
Then, for any Tarski's interpretation $I_T$ its unique extension to
all formulae $I_T^*:\L \rightarrow \mathfrak{R}$ is also the
homomorphism $I_T^*:\A_{FOL} \rightarrow \A_{\mathfrak{R}}$ from the
free syntax FOL algebra into this extensional relational algebra.\\
 Consequently, we obtain the following Intensional/extensional FOL semantics
 \cite{Majk09FOL}:\\
For any Tarski's interpretation $I_T$ of the FOL, the following
 diagram of homomorphisms commutes,
\begin{diagram}
   && &     \A_{int}~ (concepts/meaning) && &\\
  &\ruTo^{intensional~interpret.~I} && \frac{Frege/Russell}{semantics}  &&\rdTo^{h ~(extensionalization)} &\\
% & \ruTo^{I ~(intensional~int.)} & \frac{Frege/Russell}{semantics}  &\rdTo^{h (extensionaliz.)} &\\
 \A_{FOL}~(syntax)~~~~~~~~      && &\rTo_{I_T^*~(Tarski's ~interpretation)}& && ~~~~~~~~\A_{\mathfrak{R}} ~(denotation)   \\
\end{diagram}
where $h = is(w)$ where $w = I_T \in \W$ is the explicit possible
world (extensional Tarski's interpretation).\\
This homomorphic diagram formally express the fusion of Frege's and
Russell's semantics \cite{Freg92,Russe05,WhRus10} of meaning and
denotation of the FOL language, and renders mathematically correct
the definition of what we call an "intuitive notion of
intensionality", in terms of which a language is intensional if
denotation is distinguished from sense: that is, if both a
denotation and sense is ascribed to its expressions. This notion is
simply adopted from Frege's contribution (without its infinite
sense-hierarchy, avoided by Russell's approach where there is only
one meaning relation, one fundamental relation between words and
things, here represented by one fixed intensional interpretation
$I$), where the sense contains mode of presentation (here described
algebraically as an algebra of concepts (intensions) $\A_{int}$, and
where sense determines denotation for any given extensionalization
function $h$ (correspondent to a given Traski's interpretaion
$I_T$). More about the relationships between Frege's and Russell's
theories of meaning may be found in the Chapter 7,
"Extensionality and Meaning", in \cite{Beal82}.\\
As noted by Gottlob Frege and Rudolf Carnap (he uses terms
Intension/extension in the place of Frege's terms sense/denotation
\cite{Carn47}), the two logic formulae with the same denotation
(i.e., the same extension for a given Tarski's interpretation $I_T$)
need not have the same sense (intension), thus such co-denotational
expressions are not
\emph{substitutable} in general.\\
In fact there is exactly \emph{one} sense (meaning) of a given logic
formula in $\L$, defined by the uniquely fixed intensional
interpretation $I$, and \emph{a set} of possible denotations
(extensions) each determined by a given Tarski's interpretation of
the FOL as follows from Definition \ref{def:intensemant},
\begin{center}
$~~~ \L ~\rTo^I \D ~\Longrightarrow_{h = is(I_T), I_T \in ~\W }~
\mathfrak{R}$.
\end{center}
Often 'intension' has been used exclusively in connection with
possible worlds semantics, however, here we use (as many others; as
Bealer for example) 'intension' in a more wide sense, that is as an
\emph{algebraic expression} in the intensional algebra of meanings
(concepts) $\A_{int}$ which represents the structural composition of
more complex concepts (meanings) from the given set of atomic
meanings. Consequently, not only the denotation (extension) is
compositional, but also the meaning (intension) is compositional.
%----------------------------------------------------------------------
\section{Canonical models for IRDBs}
The application of the intensional FOL semantics to the Data
Integration system $\I = (\A,\S,\M)$ in Definition \ref{def:parsing}
with the user defined RDB schema $\A = (S_A, \Sigma_A)$ and the
vector big table $r_V$ can be summarized in what follows:
\begin{itemize}
  \item Each relational name (symbol) $r_k \in S_A = \{r_1,...,r_n\}$ with the arity $m =
  ar(r_k)$, is an intensional m-ary concept, so that $r_k = I(\lessdot
  r_k(\textbf{x})\gtrdot_{\textbf{x}}) \in D_m$, for a tuple of
  variables $\textbf{x} = \langle x_1,...,x_m \rangle$ and any intensional interpretation $I$.\\
  For a given Tarski's interpretation $I_T$, the extensionalization
  function $h$ is determined by $h(r_k) = \|r_k\| = \{\langle d_1,...,d_m
  \rangle \in \D^m~|~I_T(r_k(d_1,...,d_m)) = t \} = I_T(r_k) \in A$.
  The instance database $A$ of the user-defined RDB schema $\A$ is a
  model of $\A$ if it satisfies all integrity constraints in
  $\Sigma_A$.\\
  \item The relational symbol (name) $r_V$ of the vector big table
  is a particular (extensional entity), $r_V \in D_{-1}$, so that $h(r_V) = r_V$ (the name of the database $\A$). For a given model
%is an intensional entity, $r_V \in D_4$. For a given model
  $A = \{\|r_1\|,...,\|r_n\| \}$ of the user-defined RDB schema $\A$, correspondent to a given Tarski's interpretation $I_T$,
  its extension is determined by $I_T(r_V) = \|r_V\| =
  \overrightarrow{A}$.\\
  \item Intensional nature of the IRDB is evident in the fact that
  each tuple $\langle r_k,Hash(d_1,...,\\d_m),nr_{r_k}(i),d_i
  \rangle \in \overrightarrow{A}$, corresponding to the atom
  $r_V(y_1,y_2,y_3,y_4)/g$
%(with $I(y_1) = \verb"r-name" \in D_{-1},
% I(y_3) = \verb"a-name" \in D_{-1}, I(y_2) = \verb"t-index" \in D_{-1}$ and $I(y_4) = \verb"value" \in D_{-1}$)
 for an assignment $g$ such that $g(y_1) =
 r_k \in D_m,  g(y_3) =
  nr_{r_k}(i) \in D_{-1}, g(y_2) = Hash(d_1,...,d_m)) \in D_{-1}$ and $g(y_4) = d_i \in \D$, is equal to
  the intensional  tuple $\langle I(\lessdot r_k(\textbf{x}) \gtrdot_{\textbf{x}},Hash(d_1,...,d_m),nr_{r_k}(i),d_i
  \rangle$.\\ Notice that the intensional tuples are different from
  ordinary tuples composed by only particulars (extensional
  elements) in $D_{-1}$, what is the characteristics of the standard
  FOL (where the domain  of values is equal to $D_{-1}$), while here
  the "value" $r_k = I(\lessdot r_k(\textbf{x}) \gtrdot_{\textbf{x}}) \in D_m$
  is an m-ary intensional concept, for which $h(r_k) \neq r_k$ is an m-ary
  relation (while for all ordinary values $d \in D_{-1}$, $h(d) =
  d$).
  \end{itemize}
The intensional Data Integration system $\I = (\A,\S,\M)$ in
Definition \ref{def:parsing} is used in the way that the global
schema is only virtual (empty) database with a user-defined schema
$\A = (S_A, \Sigma_A)$ used to define the SQL user-defined query
which then has to be equivalently rewritten over the vector relation
$r_V$ in order to obtain the answer to this query. Thus, the
information of the database is stored only in the big table
$\|r_V\|$. Thus, the materialization of the original user-defined
schema $\A$ can be obtained by the following operation:
\begin{definition} \textsc{Materialization of the RDB}
\label{def:matz} \\Given a user-defined RDB schema $\A = (S_A,
\Sigma_A)$ with $S_A = \{r_1,...,r_n\}$ and a big vector table
$\|r_V\|$, the non SQL operation \emph{\underline{MATTER}} which
materializes the schema $\A$ into its instance database $A =
\{R_1,...,R_n\}$ where $R_k = \|r_k\|$, for $k = 1,...,n$, is given
by the following mapping, for
any  $R \subseteq \|r_V\|$:\\
$(r_k, R) ~~~\mapsto ~~~ \{\langle v_1,...,v_{ar(r_k)}\rangle ~|~
\exists y \in \pi_2(R) ((r_V(r_k,y,nr_{r_k}(1),v_1)
~\underline{\vee}~ v_1 \emph{NULL})\wedge ...\\...\wedge
(r_V(r_k,y,nr_{r_k}(ar(r_k)),v_{ar(r_k)})~\underline{\vee}~
v_{ar(r_k)} \emph{NULL})) \}$,\\
 so that the materialization of the schema $\A$ is defined by\\
   $R_k =\|r_k\| \triangleq \emph{\underline{MATTER}}(r_k, \|r_V\|)$ for each $r_k \in S_A$.
\end{definition}
The canonical models of the intensional Data Integration system $\I
= (\A,\S,\M)$ in Definition \ref{def:parsing} are the instances $A$
of the schema $\A$ such that\\
 $\|r_k\| =$ \underline{MATTER}$(r_k,
\bigcup_{\textbf{v}\in \|r_k\|}$
\underline{PARSE}$(r_k,\textbf{v}))$, that is,
when\\
$A = \{$\underline{MATTER}$(r_k, \overrightarrow{A}) ~|~r_k \in S_A\}$.\\
The canonical models of such intensional Data Integration system $\I
= \langle \A,\S,\M \rangle$ can be provided in a usual logical
framework as well:
\begin{propo} \label{prop:canonic}
Let the IRDB be given by a Data Integration system $\I = \langle
\A,\S,\M \rangle$ for a used-defined global  schema $\A =
(S_A,\Sigma_A)$ with $S_A =\{r_1,...,r_n\}$, the source schema $\S =
(\{r_V\},\emptyset)$ with the vector big data relation $r_V$ and the
set of mapping tgds $\M$ from the source schema into  he relations
of the global schema. Then  a canonical model of $\I$ is any model
of the schema $\A^+ = (S_A \bigcup \{r_V\}, \Sigma_A \bigcup \M
\bigcup \M^{OP})$, where $\M^{OP}$ is an opposite mapping tgds  from
$\A$ into $r_V$  given by the following set of tgds:\\
$\M^{OP} = \{\forall x_1,...,x_{ar(r_k)}((r_k(x_1,...,x_{ar(r_k)})
\wedge x_i \emph{NOT NULL}) \Rightarrow\\
r_V(r_k, Hash(x_1,...,x_{ar(r_k)}),nr_{r_k}(i),x_i))~|~1\leq i \leq
ar(r_k), r_k \in S_A \}$.
\end{propo}
\textbf{Proof}: It is enough to show that for each $r_k \in S_A$, and
$\textbf{x} = (x_1,...,x_{ar(r_k)})$,\\
$r_k(\textbf{x}) \Leftrightarrow
((r_V(r_k,Hash(\textbf{x}),nr_{r_k}(1),x_1) ~\underline{\vee}~ x_1
$NULL$) \wedge...\\ \wedge
(r_V(r_k,Hash(\textbf{x}),nr_{r_k}(ar(r_k)),x_{ar(r_k)})
~\underline{\vee}~ x_{ar(r_k)} $NULL$) )$.\\ From $\M^{OP}$ we have\\
 $\neg r_k(\textbf{x}) \vee \neg x_i $NOT NULL$ ~\vee~
r_V(r_k,Hash(\textbf{x}),nr_{r_k}(i),x_i)$, that is,\\
(a) $~~~\neg r_k(\textbf{x}) \vee (x_i $NULL$ ~\vee~
r_V(r_k,Hash(\textbf{x}),nr_{r_k}(i),x_i))$.\\
 From the other side,
from the fact that we have the  constraint NOT NULL for the
attribute \verb"value" (in Definition \ref{def:parsing}), then\\
 $~~~\neg r_k(\textbf{x}) \vee \neg x_i $NULL$ ~\vee~ \neg
r_V(r_k,Hash(\textbf{x}),nr_{r_k}(i),x_i))$, that is\\
(b) $~~~\neg r_k(\textbf{x}) \vee \neg (x_i $NULL$ ~\wedge~
r_V(r_k,Hash(\textbf{x}),nr_{r_k}(i),x_i))$, \\
is true and also the conjunction of (a) and (b) has to be true,
i.e.,\\ $(\neg r_k(\textbf{x}) \vee (x_i $NULL$ ~\vee~
r_V(r_k,Hash(\textbf{x}),nr_{r_k}(i),x_i))) \wedge (\neg
r_k(\textbf{x}) \vee \neg (x_i $NULL$ ~\wedge~\\ ~
r_V(r_k,Hash(\textbf{x}),nr_{r_k}(i),x_i)))$, thus, by distributivity,\\
(b) $~~~\neg r_k(\textbf{x}) \vee (x_i $NULL$ ~\underline{\vee}~
r_V(r_k,Hash(\textbf{x}),nr_{r_k}(i),x_i))$.\\
If we repeat this for all $1\leq i \leq ar(r_k)$ and conjugate all
these true formula, again by distributive property of conjunction
$\wedge$, we obtain\\
$~~~\neg r_k(\textbf{x}) \vee ((x_1 $NULL$ ~\underline{\vee}~
r_V(r_k,Hash(\textbf{x}),nr_{r_k}(1),x_1) \wedge ...\\\wedge
(x_{ar(r_k)} $NULL$ ~\underline{\vee}~
r_V(r_k,Hash(\textbf{x}),nr_{r_k}(ar(r_k)),x_{ar(r_k)}))$, that
is,\\
(c) $~~~r_k(\textbf{x}) \Rightarrow
((r_V(r_k,Hash(\textbf{x}),nr_{r_k}(1),x_1) ~\underline{\vee}~ x_1
$NULL$) \wedge...\\ \wedge
(r_V(r_k,Hash(\textbf{x}),nr_{r_k}(ar(r_k)),x_{ar(r_k)})
~\underline{\vee}~ x_{ar(r_k)} $NULL$) )$.\\
Moreover, from Definition  we also have\\
 (d) $~~~r_k(\textbf{x}) \Leftarrow
((r_V(r_k,Hash(\textbf{x}),nr_{r_k}(1),x_1) ~\underline{\vee}~ x_1
$NULL$) \wedge...\\ \wedge
(r_V(r_k,Hash(\textbf{x}),nr_{r_k}(ar(r_k)),x_{ar(r_k)})
~\underline{\vee}~ x_{ar(r_k)} $NULL$) )$,\\
That is, the logical equivalence of the formula on the left and on
the right side of the logical implication, and hence if the atom
$r_k(\textbf{x})$ is true for some assignment  to the variables $g$
so that the tuple $\langle g(x_1),...,g(x_{ar(r_k)})\rangle$ is in
relation $r_k \in \A$, then for the same assignment $g$ the every
formula\\ $r_V(r_k,Hash(\textbf{x}),nr_{r_k}(1),x_1)
~\underline{\vee}~ x_1 $NULL$,\\
...,\\ r_V(r_k,Hash(\textbf{x}),nr_{r_k}(ar(r_k)),x_{ar(r_k)})
~\underline{\vee}~ x_{ar(r_k)} $NULL \\has to be true and hence
generates the tuples in $r_V$ for  NOT NULL values of $g(x_i)$,
$1\leq i \leq ar(r_k)$.\\
Notice that the implication (c) corresponds to the nonSQL operation
\underline{PARSE}, while the implication (d) is the logical
semantics of the non SQL operation \underline{MATTER}.\\
Consequently, we obtain $\|r_k\| =$ \underline{MATTER}$(r_k,
\bigcup_{\textbf{v}\in \|r_k\|}$
\underline{PARSE}$(r_k,\textbf{v}))$, that is,
 $A = \{$\underline{MATTER}$(r_k, \overrightarrow{A}) ~|~r_k \in S_A\}$ and the database instance
 $A$ which satisfies all integrity constraints $\Sigma_A \bigcup \M
\bigcup \M^{OP}$ is the canonical model of the intensional Data
Integration system $\I = (\A,\S,\M)$.
\\$\square$\\
By joking with the words, we can say that "by PARSEing the MATTER we
obtain the pure energy" of the big vector relation, and conversely,
"by condensing the PARSEd energy we obtain the common MATTER" in a standard RDB.\\
The fact is that we do not need both of them because they are
equivalent, so instead of the more (schema) rigid RDB matter in $\A$
we prefer to use the non rigid pure energy of the big vector table
$r_V$. But we are also able to render more flexible this approach
and to decide only a subset of relations to be the intensional
concepts whose extension has to be parsed in to the vector big table
$r_V$. For standard legacy systems we can chose to avoid at all to
have the intensional concepts, thus to have the standard RDBs with
standard FOL Tarski's semantics. By declaring any of the relational
names $r_k \in S_A$ as an intensional concept, we conservatively
extend the Tarski's semantics for the FOL in order to obtain a more
expressive intensional FOL semantics for the IRDBs.\\
The fact that we assumed $r_V$ to be only a particular (extensional
entity) is based on the fact that it always will be materialized
(thus non empty relational table) as standard tables in the RDBs.
The other reason is that the extension $h(r_V)$ has not to be equal
to the vector relation (the set of tuples) $\|r_V\|$   but to the
\emph{set of relations} in the instance database $A$. Consequently,
we do not use the $r_V$ (equal to the name of the database $\A$) as
a value in the tuples of other relations and we do not use the
parsing used for all relations in the user-defined RDB
schema $\A$ assumed to be the intensional concepts as well.\\
If we would decide to use also $r_V$ as an intensional concept in
$D_4$ we would be able to parse it as all other intensional concepts
in $\A$ into itself, and such recursive definition will render (only
theoretically) an infinite extension of the $r_V$, thus non
applicable, as follows. Let $r_k = Person \in \A$ be an user-defined
relational table, and $Pname$ an attribute of this table, and let
$ID$ be the t-index value obtained by Hash function from one tuple
of $r_k$ where the value of the attribute $Pname$ is "Marco
Aurelio", then we will
have this tuples in the vector table  with (database) \emph{name} $r_V  \in D_4$:\\
1. $\langle Person, ID, Pname,  Marco Aurelio\rangle \in \|r_V\|$;\\
then by parsing this tuple 1, we will obtain for $ID_1 =
Hash(Person, ID,Pname, \\
Marco Aurelio)$ the following new tuples\\
2. $\langle r_V, ID_1, \verb"r-name",  Person \rangle \in \|r_V\|$;\\
3. $\langle r_V,  ID_1,\verb"t-index",  ID \rangle \in \|r_V\|$;\\
4. $\langle r_V,  ID_1,\verb"a-name",  Pname \rangle \in \|r_V\|$;\\
5. $\langle r_V, ID_1, \verb"value",  Marco Aurelio \rangle \in \|r_V\|$;\\
then by parsing this tuple 2, we will obtain for $ID_2 =
Hash(r_V, ID_1, \verb"r-name",\\ Person)$ the following new tuples\\
6. $\langle r_V, ID_2,  \verb"r-name", \verb"Vector" \rangle \in \|r_V\|$;\\
7. $\langle r_V, ID_2,  \verb"t-index",  ID_1 \rangle \in \|r_V\|$;\\
8. $\langle r_V,  ID_2,  \verb"a-name",  \verb"r-name" \rangle \in \|r_V\|$;\\
9. $\langle r_V, ID_2,  \verb"value",  Person \rangle \in \|r_V\|$;\\
then by parsing this tuple 2, we will obtain for $ID_3 =
Hash(r_V, ID_2,  \verb"r-name", \\Person)$ the following new tuples\\
10. $\langle r_V,  ID_3, \verb"r-name",  r_V \rangle \in \|r_V\|$;\\
...\\
Thus, as we see, the tuple 10 is equal to the tuple 6, but only with
new t-index (tuple index) value, so by continuing this process,
theoretically (if we do not pose the limits for the values of
t-indexes) we obtain an infinite process and an infinite extension
of $r_V$. Obviously, it can not happen in real RDBs, because the
length of the attribute \verb"t-index" is finite so that at some
point we will  obtained the previously generated value for this
attribute (we reach a fixed point), and from the fact that this
attribute is a part of the primary key this tuple would not be
inserted in $r_V$ because $r_V$ contains the same tuple already.\\
Notice that this process is analogous to the selfreferencing
process, where we try to use an intensional concept \emph{as an
element of itself} that has to be avoided and hence there is no
sense to render $r_V$ an intensional concept. Consequently, the IRDB
has at least one relational table which is not an intensional
concept and which will not be parsed: the vector big table, which
has this singular built-in property in every IRDB.
%----------------------------------------------------------------------
\section{NewSQL property of the IRDBs \label{sec:newSQL}}
This last section we will dedicate to demonstrate that the IRDBs are
complete w.r.t. the standard SQL. This demonstration is based on the
fact that each SQL query, defined over the user-defined schema $\A$,
which (in full intensional immersion) is composed by the intensional
concepts, will be executed over standard relational tables that
\emph{are not} the intensional concepts. If a query is defined over
the non-intensional concepts (relations) in $\A$ in this case it
will be directly executed over these relational tables as in every
RDB. If a query is defined over the intensional concepts in $\A$
(which will remain \emph{empty tables}, i.e., non materialized) then
we need to demonstrate the existence of an effective query-rewriting
 into an equivalent SQL query over the (non-intensional concept)
 vector big table $r_V$. In order to define this query-rewriting, we
 will shortly introduce the abstract syntax and semantics of
 Codd's relational algebra, as follows.\\
 Five primitive operators of Codd's algebra are: the selection, the
projection, the Cartesian product (also called the cross-product or
cross-join), the set union, and the set difference. Another
operator, rename, was not noted by Codd, but the need for it is
shown by the inventors of  Information Systems Base Language (ISBL)
for one of the earliest database management systems which
implemented Codd's relational model of data. These six operators are
fundamental in the sense that if we omit any one of them, we will
lose expressive power. Many other operators have been defined in
terms of these six. Among the most important are set intersection,
division, and the natural join. In fact, ISBL made a compelling case
for replacing the Cartesian product with the natural join, of which
the Cartesian
product is a degenerate case. \\
Recall that two relations $r_1$ and $r_2$ are union-compatible iff
$\{atr(r_1)\} = \{atr(r_2)\}$, where for a given list (or tuple) of
the attributes  $~~\textbf{a} = atr(r) = <a_1,...,a_k> =
<atr_r(1),...,atr_r(k)>$, we denote the \emph{set} $\{a_1,...,a_k\}$
by $\{atr(r)\}$, $k = ar(r)$, with the injective function
$nr_r:\{1,...,k\} \rightarrow SN$ which assigns distinct names to
each column of this relation. If a relation $r_2$ is obtained from a
given relation $r_1$ by permutating its columns, then we tell that
they are not equal (in set theoretic sense) but that they are
equivalent. Notice that in the RDB theory the two equivalent
relations are considered equal as well. In what follows, given any
two lists (tuples), $\textbf{d} = <d_1,...,d_k>$ and $\textbf{b} =
<b_1,...,b_m>$ their concatenation $<d_1,...,d_k,b_1,...,b_m>$ is
denoted by $\textbf{d} \& \textbf{b}$, where $'\&'$ is the symbol
for concatenation of the lists. By $\|r\|$ we denote the extension
of a given relation (relational symbol) $r$; it is extended to any
term $t_R$ of Codd's algebra, so that $\|t_R\|$ is the relation
obtained by computation of this term.
 Let us briefly define these
basic operators, and their correspondence with the formulae of FOL:
\index{First-Order Logic (FOL)}
\begin{enumerate}
\item Rename is a unary operation written as $\_~$ RENAME $~name_1~$AS$~name_2 $
  where the result is identical to input argument (relation) $r$ except that the
  column $i$ with  name $nr_r(i) = name_1$ in all tuples is renamed to $nr_r(i) =
  name_2$.\\
  This operation is neutral w.r.t. the logic, where we are using the
  variables for the columns of relational tables and not their
  names.
  \item Cartesian product is a binary operation $\_~ $TIMES $\_~$, written also as $\_~ \bigotimes \_~$,
   such that for the relations $r_1$ and $r_2$, first we do the rename normalization
   of $r_2$ (w.r.t. $r_1$), denoted by $r^\rho_2$, such that:\\
   For each k-th copy of the attribute $a_i$ (or, equivalently, $a_i(0)$)
    of the m-th column of $r_2$ (with $1\leq m \leq ar(r_2)$), denoted by $a_i(k) = atr_{r_2}(m) \in
   atr(r_2)$, such that the maximum index of the same attribute
   $a_i$ in $r_1$ is $a_i(n)$, we change $r_2$ by:\\
   1. $a_i(k) \mapsto a_i(k+n)$;\\
   2. if $name_1 = nr_{r_2}(m)$ is a name that exists in the set of the column
   names in $r_1$, then we change the naming function
   $nr_{r_2}:\{1,...,ar(r_2)\} \rightarrow SN$, by $nr_{r_2}(m) = name_2$, where
   $name_2 \in SN$ is a new name distinct from all other used
   names, and we define the renaming normalization $\rho$ by mapping $name_1 \mapsto name_2$.\\
   The  relation obtained from $r_2$, after this renaming
   normalization,
   will be denoted by $r^\rho_2$.
   Then we define the new relation $r$ (when both $\|r_1\| \neq \{<>\}$ and
    $\|r_2\|\neq \{<>\}$, i.e., when are not empty relations)
   by
$~  r_1 \bigotimes r^\rho_2$,\\
with  $\|r \| \triangleq \{\textbf{d}_1\& \textbf{d}_2
~|~\textbf{d}_1 \in \|r_1\|, \textbf{d}_2 \in \|r_2\|\}$, with the
naming function $nr_r:\{1,...,ar(r_1)+ar(r_2)\} \rightarrow SN$,
such that $nr_{r}(i) = nr_{r_1}(i)$ for $1\leq i \leq ar(r_1)$ and
$nr_{r}(i) = nr_{r_2}(i)$ for $1+ar(r_1)\leq i \leq
ar(r_1)+ar(r_2)$, and $atr_r:\{1,....,ar(r_1)+ar(r_2)\} \rightarrow
\textbf{att}$ function defined by $atr_{r}(i) = atr_{r_1}(i)$ for
$1\leq i \leq ar(r_1)$ and $atr_{r}(i) = atr_{r_2}(i)$ for
$1+ar(r_1)\leq i \leq
ar(r_1)+ar(r_2)$.\\
This Cartesian product is given by the following
   logical equivalence, by considering the relational symbols as predicates,\\
$r(x_1,...,x_{ar(r_1)},y_1,...,y_{ar(r_2)})\Leftrightarrow
(r_1(x_1,...,x_{ar(r_1)}) \wedge r_2(y_1,...,y_{ar(r_2)}))$, so that
  $\|r \| = \|r_1(x_1,...,x_{ar(r_1)}) \wedge r_2(y_1,...,y_{ar(r_2)})
  \|$.\\
 (if $\|r_1\|$ is empty then $~  r_1 \bigotimes r^\rho_2 = r_2$;
  if $\|r_2\|$ is empty then $~  r_1 \bigotimes r^\rho_2 = r_1$).
  \item Projection is a unary operation written as $\_~[S]$, where
  $S$ is a tuple of column names such that for a relation $r_1$
  and $S = <nr_{r_1}(i_1),...,nr_{r_1}(i_k)>$, with $k \geq 1$ and
  $1 \leq i_m \leq ar(r_1)$ for $1 \leq m \leq k$, and $i_m \neq i_j$ if $m \neq j$, we define the relation $r$ by:
$~r_1[S]$,\\
with $\|r\| =  \|r_1\|$ if $\exists name \in S.name \notin nr(r_1)$;
otherwise $\|r\| =\pi_{<i_1,...,i_k>}(\|r_1\|)$, where $nr_r(m) =
nr_{r_1}(i_m)$, $atr_r(m) = atr_{r_1}(i_m)$, for
$1 \leq m \leq k$.\\
 This projection is  given by the following
   logical equivalence\\
   $r(x_{i_1},...,x_{i_k}) \Leftrightarrow \exists
x_{j_1}...x_{j_n}r_1(x_1,...,x_{ar(r_1)})$,\\ where $n = ar(r_1)-k$
and for all $1 \leq m \leq n$, $j_m \notin \{ i_1,...,i_k\}$, so that\\
  $\|r \| = \|\exists x_{j_1}...x_{j_n}r_1(x_1,...,x_{ar(r_1)}) \|$.
  \item Selection is a unary operation written as $\_~ $WHERE$ ~C$,  where  a condition $C$ is a finite-length logical formula that consists of atoms
  %as allowed in the normal   selection
  $'(name_i ~\theta ~name_j)'~$ or $~'(name_i ~\theta ~\overline{d})~'$,
  with built-in predicates $\theta \in \Sigma_\theta \supseteq \{ \doteq,>,< \}$,  a constant $\overline{d}'$,
    and the logical operators  $\wedge$ (AND),  $\vee$ (OR) and  $\neg$
  (NOT), such that for a relation $r_1$ and $name_i$, $name_j$ the names of its columns, we define the relation $r$ by\\
  $ r_1  ~$WHERE$~ C$,\\
  as the relation with $atr(r) = atr(r_1)$ and the function $nr_r$ equal to $nr_{r_1}$, where $\|r\|$ is composed
  by the tuples in  $\|r_1\|$ for which $C$ is satisfied.\\
  This selection is  given by the following
   logical equivalence:\\
   $r(x_{i_1},...,x_{i_k}) \Leftrightarrow
  (r_1(x_1,..., x_{ar(r_1)}) \wedge C(\textbf{x}))$,\\
  where $C(\textbf{x})$ is obtained by
  substitution of each $name_i = nr_{r_1}(j)$ (of the j-th column of $r_1$)
  in the formula $C$ by the variable $x_j$, so that\\
  $\|r \| = \|r_1(x_1,..., x_{ar(r_1)}) \wedge C(\textbf{x}) \|$.\\
  4.1 We assume as an \emph{identity unary operation}, the operation $\_~ $WHERE$
  ~C$ when $C$ is the atomic condition $\overline{1} \doteq \overline{1}$
  (i.e., a tautology).
    \item Union  is a binary operation written as $\_~$ UNION $\_~$,  such that for two
    union-compatible relations $r_1 $ and $r_2$, we define the relation $r$ by:
$r_1~$ UNION $~r_2$,\\
where $\|r\| \triangleq \|r_1\| \bigcup \|r_2\|$, with $atr(r) =
atr(r_1)$, and the functions $atr_r = atr_{r_1}$, and $nr_r =
nr_{r_1}$.
%Let $r_1, r_2$ be the predicates (relational symbols) for these two
%relations, then their
This union is  given by the following
   logical equivalence:\\
   $r(x_{1},...,x_{n}) \Leftrightarrow
(r_1(x_1,...,x_{n}) \vee r_2(x_1,...,x_{n}))$,\\
 where $n = ar(r) = ar(r_1) = ar(r_2)$, so that\\
 $\|r\| = \| r_1(x_1,...,x_{n}) \vee r_2(x_1,...,x_{n})\|$.
  \item Set difference  is a binary operation written as $\_~ $MINUS$ \_~$  such that for two
  union-compatible relations $r_1$ and $r_2$, we define the relation $r$ by:
$r_1 ~$MINUS$~ r_2$,\\
where $\|r\|  \triangleq \{\textbf{t}~|~\textbf{t}\in \|r_1\|$ such
that $\textbf{t}\notin \|r_2\|\}$, with $atr(r) = atr(r_1)$, and the
functions $atr_r = atr_{r_1}$, and
$nr_r = nr_{r_1}$.\\
Let $r_1$ and $r_2$ be the predicates (relational symbols) for these
two relations. Then their difference is  given by the following
   logical equivalence:\\
   $r(x_{1},...,x_{n}) \Leftrightarrow
(r_1(x_1,...,x_{n})  \wedge \neg r_2(x_1,...,x_{n}))$,\\
 where $n = ar(r) = ar(r_1) = ar(r_2)$ and hence\\
 $\|r\| = \| r_1(x_1,...,x_{n})  \wedge \neg r_2(x_1,...,x_{n})\|$.
\end{enumerate}
Natural join $\bowtie_S$ is a binary operator, written as $(r_1
\bowtie_S r_2)$, where $r_1$ and $r_2$ are the relations. The result
of the natural join is the set of all combinations of tuples in
$r_1$ and $r_2$ that are equal on their common attribute names. In
fact, $(r_1 \bowtie_S r_2)$ can be obtained by creating the
Cartesian product $r_1\bigotimes r_2$ and then by execution of the
Selection with the condition $C$ defined as a conjunction of atomic
formulae $(nr_{r_1}(i) = nr_{r_2}(j))$ with
$(nr_{r_1}(i),nr_{r_2}(j)) \in S$ (where $i$ and $j$ are the columns
of the same attribute in $r_1$ and $r_2$, respectively, i.e.,
satisfying $atr_{r_1}(i) = atr_{r_2}(j)$) that represents the
equality of the common attribute names of $r_1$ and $r_2$.
 The natural join is arguably one of the most
important operators since it is the relational counterpart of
logical AND. Note carefully that if the same variable appears in
each of two predicates that are linked by AND, then that variable
stands for the same thing and both appearances must always be
substituted by the same value. In particular, natural join allows
the combination of relations that are associated by a foreign key.
It can also be used to define composition of binary  relations. \\
Altogether, the operators of relational algebra have identical
expressive power to that of domain relational calculus or tuple
relational calculus. However,  relational algebra is less expressive
than first-order predicate calculus without function symbols.
Relational algebra corresponds to \emph{a subset} of FOL
\index{First-Order Logic (FOL)} (denominated \emph{relational
calculus}), namely Horn clauses without recursion and negation (or
union of conjunctive queries). Consequently, relational algebra is
essentially equivalent in expressive power to \emph{relational
calculus} (and thus FOL and queries defined in Section
\ref{sec:vector}); this result is known as Codd's theorem. However,
the negation, applied to a formula of the calculus, constructs a
formula that may be true on an infinite set of possible tuples.
%, while the difference operator ofrelational algebra always returns a finite result.
To overcome this difficulty, Codd restricted the operands of
relational algebra to finite relations only and also proposed
restricted support for negation $\neg$ (NOT) and disjunction $\vee$
(OR). Codd defined the term "relational completeness" to refer to a
language that is complete with respect to first-order predicate
calculus apart from the restrictions he proposed. In practice, the
restrictions have no adverse effect on the applicability of his
relational algebra for database purposes. \\
Several papers have proposed new operators of an algebraic nature as
candidates for addition to the original set. We choose the
additional unary operator 'EXTEND$\_~$ADD $a,name$ AS $e$' denoted
shortly as $\_~\langle a,name,e\rangle $, where $a$ is a new added
attribute as the new column (at the end of relation) with a new
fresh name $name$ and $e$ is an expression (in the most simple cases
it can be the value NULL or a constant $\overline{d}$, or the i-th
column name $nr(i)$ of the argument (i.e., relation) of this
operation), for \emph{update} relational algebra operators, in order
to cover all of the basic features of data manipulation (DML)
aspects of a relation models of data, so that
\begin{itemize}
  \item We define a unary operator $\_~\langle a,name,e\rangle $, for an attribute
  $a \in \textbf{att}$, its name, and expression $e$, as a function with a
  set of column names, such that
  for a relation $r_1$ and expression $e$ composed of the
  names of the columns of $r_1$ with $n = ar(r_1)$, we obtain the
  $(ar(r_1)+1)$-ary
  relation $r $ by
    $~ \langle a,name, e\rangle (r_1)$,\\
  with naming function $nr_r:\{ar(r_1)+1\} \rightarrow SN$ such
  that $nr_r(i) = nr_{r_1}(i)$ if $i \leq ar(r_1)$; $nr_r(ar(r_1)+1) = name$ otherwise, being a
  fresh new name for this column; with the attribute function $atr_r:\{ar(r_1)+1\} \rightarrow \textbf{att}$ such
  that $atr_r(i) = atr_{r_1}(i)$ if $i \leq ar(r_1)$; $atr_r(ar(r_1)+1) = a$ otherwise, and
 \\ $\|r\| = \{<>\} \bigcup \{\textbf{d}\& e(\textbf{d})~|~\textbf{d} \in \|r_1\|
 \}$,\\
   where $e(\textbf{d}) \in dom(a)$ is a constant or the value obtained from the function
   $e$ where each name $nr_r(i)$ is substituted by the value $d_i$
   of the tuple $\textbf{d} = <d_1,...,d_n> \in \|r_1\|$; in the special
   cases, we can use nullary functions (constants) for the expression
   $e$ (for example, for the NULL value).\\
   (note that $r$ is empty if $e$ is an expression and $r_1$ empty
   as well).\\
   Then, for a nonempty relation $r_1$, the EXTEND~$ r_1$ ADD $a,name$ AS $e$ (i.e., $r_1 \langle a,name,e\rangle
   $)
   %is derived from it and    equal to  \\$~(r_1$
   %TIMES ($\langle a,name, e\rangle (r_1)$)) WHERE $(name = e)$ if $ar(r_1)$ is finite; $r_1$ otherwise.\\
   %This composed algebraic operations
    can be represented by the following
   logical equivalence:\\
   $~r(x_1,...,x_{n+1}) \Leftrightarrow (r_1(x_1,...,x_{n}) \wedge (x_{n+1} =
   e(\textbf{x})))$\\
   where $e(\textbf{x})$ is obtained by
  substituting each $name_i = nr_{r_1}(j)$ (of the j-th column
   of $r_1$) in the expression $e$ by the variable $x_j$.
   \end{itemize}
We are able to define a new relation with a single tuple $\langle
\overline{d}_1,..,\overline{d}_k \rangle, k \geq 1$ with the given
list of attributes $\langle a_1,..,a_k \rangle$, by the following
finite length expression,\\
EXTEND (...(EXTEND $r_\emptyset$ ADD $a_1,name_1$ AS
$\overline{d}_1)...)$ ADD $a_k,name_k$ AS $\overline{d}_k$, or
equivalently by
$r_\emptyset \langle a_1,name_1, \overline{d}_1\rangle  \bigotimes...\bigotimes r_\emptyset\langle a_k,name_k, \overline{d}_k\rangle$,\\
 where $r_\emptyset$ is the empty
type relation  with $\|r_\emptyset \|= \{<>\}$, $ar(r_\emptyset) =
0$ introduced in Definition \ref{def:bealer}, and empty functions
$atr_{r_\emptyset}$ and $nr_{r_\emptyset}$. Such single tuple
relations can be used for an insertion in a given relation (with the
same list of attributes) in
what follows.\\
 \textbf{Update operators.} The three update operators,
'UPDATE', 'DELETE' and 'INSERT'  of the Relational algebra, are
derived operators from these previously defined operators  in the
following way:
\begin{enumerate}
  \item Each algebraic formulae 'DELETE FROM $~ r$ WHERE C' is
equivalent to the formula '$r$ MINUS ($r$ WHERE $C$)'.
  \item Each algebraic expression (a term) 'INSERT INTO $~r[S]$ VALUES (list of values)', 'INSERT INTO $~r[S]$ AS SELECT...',
%where $r,   r_1$ are two union compatible relations, or the two algebraic expressions which have as result
%two union compatible relations,
is equivalent to  '$r~$ UNION $~r_1$' where the union compatible
relation $r_1$ is a one-tuple relation
(defined by list) in the first,  or a relation defined by 'SELECT...' in the second case.\\
In the case of a single tuple insertion (version with 'VALUES') into
a given relation $r$, we can define a single tuple relation $r_1$ by
using 'EXTEND..' operations.
  \item Each algebraic expression 'UPDATE $~r$ SET $[nr_r(i_1)= e_{i_1} ,..., nr_r(i_k) =
e_{i_k}]$ WHERE  $C$', for  $n =ar(r)$, where $e_{i_m}, 1 \leq i_m
\leq n$ for $1 \leq m\leq k$ are the expressions and $C$ is a
condition, is equal to the formula '($r$
WHERE  $\neg C$) UNION $r_1$' , where $r_1$ is a relation expressed by\\
  (EXTEND(...(EXTEND ($r$ WHERE $C$) ADD $att_r(1),name_1$ AS $e_1$)...) ADD $att_r(n),name_n$ AS
 $e_{n})[S]$,\\
 such that for each $1 \leq m \leq n$, if $m \notin \{i_1,...,i_k \}$ then $e_m  =
 nr_r(m)$, and $S = <name_1,...,name_n>$.
\end{enumerate}
Consequently, all update operators of the relational algebra can be
obtained by addition of these 'EXTEND $\_~$ ADD $a,name$ AS $e$'
operations. \\
Let us define the $\Sigma_R$-algebras sa follows (\cite{Majk14},
Definition 31 in Section 5.1):
\begin{definition} \label{def:relAlg}
  We denote the algebra of the set of operations, introduced previously in
this section (points from 1 to 6 and \emph{EXTEND} $\_~$ \emph{ADD}
$a,name$ \emph{AS} $e$) with additional nullary operator
(empty-relation constant) $\perp$, by $\Sigma_{RE}$. Its subalgebra
without $\_~ $\emph{MINUS}$ \_~$ operator is denoted by
$\Sigma_R^+$, and without $\perp$ and unary operators \emph{EXTEND}
$\_~$ \emph{ADD} $a,name$ \emph{AS} $e$  is denoted by $\Sigma_R$
(it is the "select-project-join-rename+union" (\emph{SPJRU})
subalgebra). We define the  set of terms $\T_PX$ with variables in
$X$
%(of all finite-length terms with variables in $X$)
of this $\Sigma_R$-algebra (and analogously for the terms $\T_P^+X$
of $\Sigma_R^+$-algebra),
inductively as follows:\\
1.  Each relational symbol (a variable) $r \in X \subseteq
\mathbb{R}$ and  a constant (i.e., a nullary operation) is a term in
 $\T_P X$;\\
2. Given any term $t_R \in \T_P X$ and an unary operation $o_i \in
\Sigma_R$, $o_i(t_R)\in  \T_P X$;\\
3. Given any two terms $t_R, t'_R \in \T_P X$ and a binary operation
$o_i \in \Sigma_R$, $o_i(t_R,t'_R) \in \T_P X$.\\
We define the evaluation of terms in $\T_P X$, for $X = \mathbb{R}$,
by extending the assignment  $\|\_~\|:\mathbb{R} \rightarrow
\underline{\Upsilon}$, which assigns a relation to each relational
symbol (a variable) to all terms by the function $\|\_~\|_{\#}:\T_P
\mathbb{R} \rightarrow \underline{\Upsilon}$ (with $\|r\|_{\#} =
\|r\|$), where $\underline{\Upsilon}$ is the universal database
instance (set of all relations for a given universe $\D$). For a
given term $t_R$ with relational symbols $r_1,..,r_k \in
\mathbb{R}$, $\|t_R\|_{\#}$ is the relational table obtained from
this expression for the given set of relations $\|r_1\|,...,\|r_k\|
\in \underline{\Upsilon}$, with the constraint that \\ $\|t_R $
\emph{UNION} $t'_R\|_{\#} = \|t_R \|_{\#} \bigcup \|t'_R\|_{\#}$ if
the relations $\|t_R \|_{\#}$ and $\|t'_R \|_{\#}$ are
union compatible; $\perp = \{<>\} = \|r_\emptyset\|$ (empty relation) otherwise.\\
%Each R-algebra $\alpha:X\rightarrow \underline{\Upsilon}$ is a
%restriction of an assignment $~\|\_~\|$ to $X \subseteq \mathbb{R}$.\\
We say that two terms $t_R, t'_R \in \T_P X$ are equivalent (or
equal), denoted by $t_R \approx t'_R$, if for all assignments
$~\|t_R\|_{\#} = \|t'_R\|_{\#}$.
\end{definition}
 We say
that an extension $\|t_R\|_{\#}$,  of a term $t_R \in \T_PX$, is
\emph{vector relation} of the \emph{vector view} denoted by
$\overrightarrow{t_R}$ if the type of $\|t_R\|_{\#}$ is equal to the
type of the vector relation $r_V$.\\
Let $R = \|\overrightarrow{t_R}\|_{\#}$ be the relational table with
the four attributes (as $r_V$)
\verb"r-name",\verb"t-index"\\\verb"a-name" and \verb"value", then
 its used-defined view representation can be derived as follows:
\begin{definition} \textsc{View Materialization}: \label{def:view-mater}
Let $t_R \in \T_PX$ be a user-defined SPJU
(Select-Project-Join-Union) view over a database schema $\A =
(S_A,\Sigma_A)$ with the type (the tuple of the view columns)
$\mathfrak{S} =
\langle(r_{k_1},name_{k_1}),...,(r_{k_m},name_{k_m})\rangle$, where
the $i$-th column $(r_{k_i},name_{k_i})$ is the column  with name
equal to $name_{k_i}$ of the relation name $r_{k_i}\in S_A$, $1\leq
i \leq m$, and $\overrightarrow{t_R}$ be the rewritten query over
$r_V$. Let $R = \|\overrightarrow{t_R}\|_{\#}$ be the resulting
relational table with the four attributes (as $r_V$)
\verb"r-name",\verb"t-index",\verb"a-name" and \verb"value". We
define the operation \emph{\underline{VIEW}} of the transformation
of $R$ into the user defined view
representation by:\\
$\emph{\underline{VIEW}}(\mathfrak{S},R) = \{\langle
d_1,...,d_m\rangle ~|~\exists ID \in \pi_3(R), \forall_{1\leq i \leq
m}(\langle r_{k_i},name_{k_i},ID,d_i \rangle \in R$; otherwise set
$d_i$ to \emph{NULL} $\}$.
\end{definition}
Notice that we have  $\|r_k\| =$ \underline{VIEW}$(\mathfrak{S},R) =
$\underline{MATTER}$(r_k,R)$ for each $r_k \in S_A$ with $R =
\bigcup_{\textbf{d} \in\|r_k\|}$\underline{PARSE}$(r_k,\textbf{d})$,
and $\mathfrak{S} =
\langle(r_{k},nr_{r_k}(1)),...,(r_{k},nr_{r_k}(ar(r_k)))\rangle$,
and hence the nonSQL operation \underline{MATTER} is a special case
of the operation \underline{VIEW}.\\
 For any original user-defined query (term)
$t_R$ over a user-defined database schema $\A$, by
$\overrightarrow{t_R}$ we denote the equivalent (rewritten) query
over the vector relation $r_V$. We have the following important
result for the IRDBs:
\begin{propo} \label{prop:newSQL} There exists a complete algorithm for the
 term rewriting of any user-defined SQL term $t_R$ over a schema $\A$, of the
full  relational algebra $\Sigma_{RE}$ in Definition
\ref{def:relAlg},  into an equivalent vector query
$\overrightarrow{t_R}$ over the vector relation $r_V$. \\If $~t_R$
is a SPJU  term (in Definition \ref{def:view-mater}) of the type
$\mathfrak{S}$ then $\|t_R\|_{\#} =
\emph{\underline{VIEW}}(\mathfrak{S},\|\overrightarrow{t_R}\|_{\#})$.
\end{propo}
\textbf{Proof:} In this proof we will use the convention that two
NULL values can not be compared as equal, because their meaning is
that the value is missing, and two missing values not necessarily ar
equal. In fact in $r_V$ we do not store the null values but consider
them as unknown missing values. Thus, when there are null values in
the columns of the user-defined tables being joined, the null values
do not match each other. \\Let us show that there is such a query
(relation algebra term) rewriting for each basic relational operator
previously described, recursively (in what follows, if $~t_R = r$
then $~\overrightarrow{t_R} = ~r_V$ WHERE $\verb"r-name" = r$):
\begin{enumerate}
\item (Rename).  $t'_R = r~$ RENAME $~name_1~$AS$~name_2 $
  where the result is identical to input argument (relation) $r$ except that the
  column $i$ with  name $nr_r(i) = name_1$ in all tuples is renamed to $nr_r(i) =
  name_2$.  The rewritten vector query is\\
  $\overrightarrow{t'_R}=  ~$UPDATE$ ~r_V~ $SET$~ [\verb"a-name" = name_2]$ WHERE $(\verb"r-name" = r)\wedge (\verb"a-name" =
  name_1)$;\\
  \item (Projection).  $t'_R = ~t_R~[S]$, where
    and $S = \langle (r_{j_1},nr_{r_{j_1}}(i_1)),...,(r_{j_k},nr_{r_{j_k}}(i_k))\rangle \subseteq \mathfrak{S}$, with $k \geq 1$ and
  $1 \leq i_m \leq ar(r_{j_m})$ for $1 \leq m \leq k$, is a subset  of the type $\mathfrak{S}$ of the term $t_R$. We define the rewritten vector
  query\\
$\overrightarrow{t'_R} = \overrightarrow{~t_R~[S]}\\ =
~\overrightarrow{t_R}$ WHERE $((nr_{\overrightarrow{t_R}}(1) =
r_{j_1}) \wedge (nr_{\overrightarrow{t_R}}(2) = nr_{r_{j_1}}(i_1)))
\vee ...\vee ((nr_{\overrightarrow{t_R}}(1) = r_{j_k})\\ \wedge
(nr_{\overrightarrow{t_R}}(2) = nr_{r_{j_k}}(i_k)))$;\\
  \item (Join). $t'_R = t_{R,1} \bowtie_S t_{R,2}$, where $S =
  (((r_{l_1},nr_{r_{l_1}}(i_1)),(r_{n_1},nr_{r_{n_1}}(j_1))),...,\\(((r_{l_m},nr_{r_{l_m}}(i_m)),(r_{n_m},nr_{r_{n_m}}(j_m)))))$
  with $1\leq i_k \leq |\mathfrak{S}_1|$ and $1\leq j_k \leq |\mathfrak{S}_2|$ for $1\leq k \leq
  m$, where $\mathfrak{S}_1$ and $\mathfrak{S}_2$ are the types of $t_{R,1}$ and $t_{R,2}$, respectively. .\\ Let us define the following relational algebra terms:\\
  $r = \overbrace{\overrightarrow{t_{R,1}}\bigotimes...\bigotimes \overrightarrow{t_{R,1}}}^{m}\bigotimes \overbrace{\overrightarrow{t_{R,2}}\bigotimes...\bigotimes \overrightarrow{t_{R,2}}}^{m}~$;
  (the first $m$ are for the attributes of $\overrightarrow{t_{R,1}}$ in $S$, and the last $m$
  are for the corresponded joined attributes of $\overrightarrow{t_{R,2}}$ in $S$).
Note that each column name  of $\overrightarrow{t_{R,1}}$ will be
renamed $m$ times  in order to have for each column different name
in standard way as for attributes: for example, the column
$\verb"a-name"$ in $\overrightarrow{t_{R,1}}$ will be repeated by
$\verb"a-name"(1),..., \verb"a-name"(m)$ (and similarly for each
column name of $\overrightarrow{t_{R,2}}$), and the attribute
$at_{r}(2)$ (of the column $\verb"t-index"$ in $r_V$) will be
repeated by its copies $at_{r}(2)(1),...,at_{r}(2)(2m)$ (in what
follows will be also generated  the $(2m+1)$-th copy
$at_{r}(2)(2m+1)$ in the algebra term $t_2$). Thus,
  \\
  $t_1 = r$ WHERE $
  (\bigwedge_{1\leq k \leq m} ((nr_r(4k-3) = r_{l_k}) \wedge (nr_r(4k-1) = nr_{r_{l_k}}(i_k)) \wedge (nr_r(4(m+k)-3) =
  r_{n_k})
  \wedge (nr_r(4(m+k)-1) = nr_{r_{n_k}}(j_k)) \wedge (nr_r(4k) = nr_r(4(m+k)+4)))\\
\wedge ((m =1) \vee ((nr_r(2) = nr_r(6) = ... = nr_r(4m-2)) \\
\wedge (nr_r(4m+2)
= nr_r(4m+6) = ... = nr_r(8m-2)))))$;\\
  $t_2 = ($EXTEND $t_1$ ADD $((atr_r(2))(2m+1), name_3,
  Hash(atr_r(1),atr_r(2),...,\\atr_r(8m))))[nr_{t_1}(2), nr_{t_1}(4m+2),name_3]$; \\
  where $name_3$ is a fresh new name and \\$\|t_2\|_{\#} = \{\langle ID_1,ID_2,ID_3 \rangle
  ~|~ ID_1$ is the tuple-index in $\|t_{R,1}\|_{\#}$ and $ID_2$ is the corresponding joined tuple-index in $\|t_{R,2}\|_{\#}$, while $ID_3$ is the fresh new generated (by Hash function) tuple-index  for
  the tuple obtained by join operation $\}$,\\ then for  the Cartesian products $t_3 = \overrightarrow{t_{R,1}} \bigotimes t_2$ and $t_4 = \overrightarrow{t_{R,2}} \bigotimes t_2$,\\
    $\overrightarrow{t'_R} = \overrightarrow{t_{R,1} \bowtie_S t_{R,2}} \\
    =((t_3$ WHERE $(nr_{t_3}(2) = nr_{t_3}(5)))[nr_{t_3}(1),
   name_3,nr_{t_3}(3),nr_{t_3}(4)])$\\
   UNION $((t_4$ WHERE $(nr_{t_4}(2) = nr_{t_4}(6)))[nr_{t_4}(1),
   name_3,nr_{t_4}(3),nr_{t_4}(4)])$;\\
   \item (Selection).  $t'_R = ~t_R~ $WHERE$ ~C$:\\
   4.1  When  a condition $C$ is a finite-length logical formula that consists of atoms
  $'((r_{i_1},name_i) ~\theta ~(r_{j_1},name_j))'~$ or $~'(r_{i_1},name_i) ~\theta
  ~\overline{d}~'$ or $~'(r_{i_1},name_i) ~$ NOT NULL' with built-in predicates $\theta \in \Sigma_\theta \supseteq \{ \doteq,>,< \}$,  a constant $\overline{d}'$,
    and the logical operators, between the columns in the type $\mathfrak{S}$ of the term
  $t_R$. \\The condition $C$, composed by $k \geq 1$ different columns
  in $\mathfrak{S}$, we denote by
  $C((r_{i_1},name_{i_1}),...,(r_{i_k},name_{i_k}))$, $k \geq 1$, and hence we define the rewritten vector
    query\\
  $\overrightarrow{t'_R} =  \overrightarrow{~t_R~ $WHERE$ ~C} = ~\overrightarrow{t_R}$ WHERE $nr_{\overrightarrow{t_R}}(2)$ IN $t_1$\\
  where for  $~r = \overbrace{\overrightarrow{t_R}\bigotimes ... \bigotimes
  \overrightarrow{t_R}}^k ~$  we define the unary relation which
  contains the tuple-indexes of the relation $\|t_R\|$ for its
  tuples which satisfy the selection condition $C$,
  by the following selection term \\$t_1 = (r$ WHERE $((nr_r(1) = r_{i_1} \wedge nr_r(3) =
  name_{i_1}) \wedge...\wedge (nr_r(1+4(k-1)) = r_{i_k} \wedge nr_r(3+4(k-1)) =
  name_{i_1}))  \wedge C(nr_r(4),...,nr_r(4k)) \\\wedge ((k = 1)\vee (nr_r(2) = nr_r(6) =... = nr_r(2 +4(k-1)))))
  [nr_r(2)]$;\\
  4.2 Case when $C = ~(r_{i_1},name_i) ~$ NULL,\\
$\overrightarrow{t'_R} =  \overrightarrow{~t_R~
$WHERE$~(r_{i_1},name_i) ~$ NULL $} = \overrightarrow{t_R} $ WHERE
$nr_{\overrightarrow{t_R}}(2)$ NOT IN $t_2$,\\
where $t_2 = (\overrightarrow{~t_R~} $WHERE
$((nr_{\overrightarrow{t_R}}(1) = r_{i_1}) \wedge
(nr_{\overrightarrow{t_R}}(3) =
name_i)))[(nr_{\overrightarrow{t_R}}(2)]$.\\
From the fact that $~t_R~ $WHERE $~C_1 \wedge C_2 = ~(t_R~ $WHERE
$~C_1$) WHERE $C_2$ and $~t_R~ $WHERE $~C_1 \vee C_2 = ~(t_R~ $WHERE
$~C_1$) UNION $~(t_R~ $WHERE $~C_2)$, and De Morgan laws, $\neg (C_1
\wedge C_2) = \neg C_1 \vee \neg C_2$, $\neg (C_1 \vee C_2) = \neg
C_1 \wedge \neg C_2$, we can always divide any selection in the
components of the two disjoint cases above;\\
    \item (Union).   $t'_R = t_{R,1}~$ UNION$_R ~t_{R,2}$,
where $R$ is a table $\{\langle r_{l_k},name_{l_k},
r_{n_k},name_{n_k}\rangle ~|~\\1\leq k \leq m\}$ such that
$\mathfrak{S}_1 = \langle (r_{l_1},name_{l_1}),...,
(r_{l_m},name_{l_m})\rangle $  and $\mathfrak{S}_2 = \langle
(r_{n_1},name_{n_1}),..., (r_{n_m},name_{n_m})\rangle $ are the
types of $t_{R,1}$ and $t_{R,1}$, respectively, with the
union-compatible columns $(\langle r_{l_k},name_{l_k})$ and
$(\langle r_{n_k},name_{n_k})$ for every $1\leq k \leq m$.\\  We
define the relational algebra term   $t_1 =
\overrightarrow{t_{R,2}} \bigotimes  R$, so that\\
$\overrightarrow{t'_R} = \overrightarrow{t_{R,1}~$ UNION$_R ~t_{R,2}}\\
= \overrightarrow{t_{R,1}}$ UNION $((t_1$ WHERE $((nr_{t_1}(1) =
nr_{t_1}(7)) \wedge (nr_{t_1}(3) =
nr_{t_1}(8))))[nr_{t_1}(5),\\nr_{t_1}(2),nr_{t_1}(6),nr_{t_1}(4)]),$\\
so that the relation-column names of the union will be equal to the
column names of the first term in this union;\\
\item (Set difference).   $t'_R = t_{R,1}~$ MINUS$_R ~t_{R,2}$,
where where $R$ is a table $\{\langle r_{l_k},nr_{r_{l_k}}(i_k),\\
r_{n_k},nr_{r_{n_k}}(j_k)\rangle ~|~1\leq k \leq m\}$ such that
 $\mathfrak{S}_1 = \langle (r_{l_1},nr_{r_{l_1}}(i_1)),...,
(r_{l_m},nr_{r_{l_m}}(i_m))\rangle $  and $\mathfrak{S}_2 = \langle
(r_{n_1},nr_{r_{n_1}}(j_1)),..., (r_{n_m},nr_{r_{n_m}}(j_m))\rangle
$ are the types of $t_{R,1}$ and $t_{R,1}$, respectively, with the
union-compatible columns $(\langle r_{l_k},nr_{r_{l_k}}(i_k))$ and
$(\langle r_{n_k},nr_{r_{n_k}}(j_k))$ for every $1\leq k \leq m$.
Let us define the following relational algebra terms:\\
$r = \overbrace{\overrightarrow{t_{R,1}}\bigotimes...\bigotimes
\overrightarrow{t_{R,1}}}^{m}\bigotimes
\overbrace{\overrightarrow{t_{R,2}}\bigotimes...\bigotimes
\overrightarrow{t_{R,2}}}^{m}~$;
  (the first $m$ are for the attributes of $\overrightarrow{t_{R,1}}$ in $\mathfrak{S}_1$, and the last $m$
  are for the corresponded joined attributes of $\overrightarrow{t_{R,2}}$ in $\mathfrak{S}_2$).
 Thus,  \\   $t_1 = (r$ WHERE $
  (\bigwedge_{1\leq k \leq m} ((nr_r(4k-3) = r_{l_k}) \wedge (nr_r(4k-1) = nr_{r_{l_k}}(i_k)) \wedge (nr_r(4(m+k)-3) =
  r_{n_k})
  \wedge (nr_r(4(m+k)-1) = nr_{r_{n_k}}(j_k)) \wedge (nr_r(4k) = nr_r(4(m+k)+4)))\\
\wedge ((m =1) \vee ((nr_r(2) = nr_r(6) = ... = nr_r(4m-2)) \\
\wedge (nr_r(4m+2) = nr_r(4m+6) = ... = nr_r(8m-2))))))[nr_{t_1}(2)]$;\\
    where  $\|t_1\|_{\#} = \{\langle ID_1\rangle  ~|~
  ID_1$ is the tuple-index  of a tuple in $\|t_{R,1}\|_{\#}$ for which there exists an equal tuple in
  $\|t_{R,2}\|_{\#}\}$. Then,\\
$\overrightarrow{t'_R} = \overrightarrow{t_{R,1}~$ MINUS$_R
~t_{R,2}} = \overrightarrow{t_{R,1}}$ WHERE
$nr_{\overrightarrow{t_{R,1}}}(2)$ NOT IN $t_1$.
\end{enumerate}
It is easy to show that for the cases from 2 to 6, we obtain that
$\|t'_R\|_{\#} =
\emph{\underline{VIEW}}(\mathfrak{S},\|\overrightarrow{t'_R}\|_{\#})$,
where $\mathfrak{S}$ is the type of the relational algebra term
$t'_R$. Thus, for any SPJU term $t_R$ obtained by the composition of
these basic relational algebra operators we have that $\|t_R\|_{\#}=
\emph{\underline{VIEW}}(\mathfrak{S},\|\overrightarrow{t_R}\|_{\#})$.\\
The update operators are rewritten as follows:
\begin{enumerate}
  \item (Insert).  INSERT INTO $~r[S]~$ VALUES $~(d_1,...,d_m)$, where $S
= \langle nr_r(i_1),...,\\nr_r(i_m) \rangle$, $1\leq m \leq ar(r)$,
is the subset of mutually different attribute names of $r$ and all
$v_i$, $1\leq i \leq m$ are the values different from NULL. It is
rewritten into the following set of terms:\\ $\{$ INSERT INTO
$~r_V[\verb"r-name",  \verb"t-index", \verb"a-name",\verb"value"]~$
VALUES $~(r,Hash(d_1,\\...,d_m),nr_r(i_k),d_k)~|~1
\leq k \leq m \}$.\\
Note that before the execution of this set of insertion in $r_V$,
the DBMS has to control if it satisfy all user-defined integrity
constraints in the user-defined database schema $\A$;\\
  \item (Delete). DELETE FROM $~r~$ WHERE $~C$, is rewritten into the
  term:\\
  DELETE FROM $~r_V~$ WHERE \verb"t-index" IN
  $~\overrightarrow{t_R}[nr_{\overrightarrow{t_R}}(2)]$,\\
  where $\overrightarrow{t_R} = \overrightarrow{r~$ WHERE $~C}$ is the selection term as described in point 4 above;\\
  \item (Update). the existence of the rewriting of this operation
  is  obvious, from the fact that it can always be decomposed as
  deletion and after that the insertion of the tuples.
\end{enumerate}
$\square$\\
This proposition demonstrates that the IRDB is full SQL database, so
that each user-defined query over the used-defined RDB database
schema $\A$ can be equivalently transformed by query-rewriting into
a query over the vector relation $r_V$. However, in the IRDBMSs we
can use more powerful and efficient algorithms in order to execute
each original user-defined query over the vector relation $r_V$.\\
Notice that this proposition demonstrates that the IRDB is a kind of
GAV Data Integration System $\I = (\A,\S,\M)$ in  Definition
\ref{def:parsing} where we do not materialize the user-defined
schema $\A$ but only the vector relation $r_V \in \S$ and each
original query $q(\textbf{x})$ over the empty schema $\A$ will be
rewritten into a vector query $\overrightarrow{q(\textbf{x})}$ of
the type $\mathfrak{S}$ over the vector relation $r_V$, and then the
resulting view
\underline{VIEW}$(\mathfrak{S},\|\overrightarrow{q(\textbf{x})}\|_{\#})$
will be returned to user's application.\\
Thus, an IRDB is a member of the NewSQL, that is, a member of a
class of modern relational database management systems that seek to
provide the same scalable performance of NoSQL systems for online
transaction processing (read-write) workloads while still
maintaining the ACID guarantees of a traditional database system.
%----------------------------------------------------------------------
%----------------------------------------------------------------------
% SECTION VII: Conclusions
%----------------------------------------------------------------------
\section{Conclusion}
The method of parsing of a relational instance-database $A$ with the
user-defined schema $\A$ into a vector relation
$\overrightarrow{A}$, used in order to represent the information in
a standard and simple key/value form, today in various applications
of Big Data, introduces the intensional concepts for the
user-defined relations of the schema $\A$. In Tarskian semantics of
the FOL used to define the semantics of the standard RDBs, one
defines what it takes for a sentence in a language to be true
relative to a model. This puts one in a position to define what it
takes for a sentence in a language to be valid. Tarskian semantics
often proves quite useful in logic. Despite this, Tarskian semantics
neglects meaning, as if truth in language were autonomous. Because
of that the Tarskian theory of truth becomes inessential to the
semantics for more expressive logics, or more 'natural' languages.\\
Both, Montague's and Bealer's approaches were useful for this
investigation of the intensional FOL with intensional abstraction
operator, but the first is not adequate and explains why we adopted
two-step intensional semantics (intensional interpretation with the
set of extensionalization functions). Based on this intensional
extension of the FOL, we defined a new family of IRDBs. We have
shown that also with this extended intensional semantics we may
continue to use the same SQL used for the RDBs.\\
This new family of IRDBs extends the traditional RDBS with new
features. However, it is compatible in the way how to present the
data by user-defined database schemas (as in RDBs) and with SQL for
management of such a relational data. The  structure of RDB is
parsed into a vector key/value relation so that we obtain a column
representation of data used in Big Data applications, covering the
key/value and column-based Big Data applications as well, into a
unifying RDB framework.\\
Note that the method of parsing is well suited for the migration
from  all existent RDB applications where the data is stored in the
relational tables, so that this solution gives the possibility to
pass easily from the actual RDBs into the new machine engines for
the IRDB. We preserve all metadata (RDB schema definitions) without
modification and only dematerialize its relational tables by
transferring their stored data into  the vector relation $r_V$
(possibly in a number of disjoint partitions over a number of
nodes). From the fact that we are using the query rewriting IDBMS,
the current user's (legacy) applications does not need any
modification and they continue to "see" the same user-defined RDB
schema as before. Consequently, this IRDB solution is adequate for a
massive migration from the already obsolete and slow RDBMSs into a
new family of fast, NewSQL schema-flexible (with also 'Open
schemas') and Big Data scalable IRDBMSs.

%----------------------------------------------------------------------

%\bibliographystyle{abbrv}
\bibliographystyle{IEEEbib}
\bibliography{mydb}

%\newpage
%\input{hybridtheory}
%\balancecolumns

%$\vspace*{-4mm}$

%
\end{document}